\documentclass[twocolumn,twocolappendix,trackchanges]{aastex63}

\received{August 26, 2020}
\revised{November 12, 2020}
\accepted{November 30, 2020}
\shorttitle{Long-lasting SNe from Subaru/HSC}
\shortauthors{Moriya et al.}
\graphicspath{{./}{figures/}}
\begin{document}

\title{
Constraints on the rate of supernovae lasting for more than a year from Subaru/Hyper Suprime-Cam
}

\correspondingauthor{Takashi J. Moriya}
\email{takashi.moriya@nao.ac.jp}

\author[0000-0003-1169-1954]{Takashi J. Moriya}
\affiliation{National Astronomical Observatory of Japan, National Institutes of Natural Sciences, 2-21-1 Osawa, Mitaka, Tokyo 181-8588, Japan}
\affiliation{School of Physics and Astronomy, Faculty of Science, Monash University, Clayton, Victoria 3800, Australia}

\author[0000-0002-9092-0593]{Ji-an Jiang}
\affiliation{Kavli Institute for the Physics and Mathematics of the Universe (WPI), The University of Tokyo Institutes for Advanced Study, The University of Tokyo, 5-1-5 Kashiwanoha, Kashiwa, Chiba 277-8583, Japan}
\author{Naoki Yasuda}
\affiliation{Kavli Institute for the Physics and Mathematics of the Universe (WPI), The University of Tokyo Institutes for Advanced Study, The University of Tokyo, 5-1-5 Kashiwanoha, Kashiwa, Chiba 277-8583, Japan}
\author{Mitsuru Kokubo}
\affiliation{Astronomical Institute, Tohoku University, 6-3 Aramaki Aza-Aoba, Aoba, Sendai, Miyagi 980-8578, Japan}

% HSC internal members
\author{Kojiro Kawana}
\affiliation{Department of Physics, Graduate School of Science, The University of Tokyo, 7-3-1 Hongo, Bunkyo, Tokyo 113-0033, Japan}
\author{Keiichi Maeda}
\affiliation{Department of Astronomy, Kyoto University, Kitashirakawa-Oiwake-cho, Sakyo-ku, Kyoto 606-8502, Japan}
\affiliation{Kavli Institute for the Physics and Mathematics of the Universe (WPI), The University of Tokyo Institutes for Advanced Study, The University of Tokyo, 5-1-5 Kashiwanoha, Kashiwa, Chiba 277-8583, Japan}
\author{Yen-Chen Pan}
\affiliation{Graduate Institute of Astronomy, National Central University, 300 Jhongda Road, Zhongli, Taoyuan, 32001, Taiwan}
\author{Robert M. Quimby}
\affiliation{Department of Astronomy / Mount Laguna Observatory, San Diego State University, 5500 Campanile Drive, San Diego, CA, 92812-1221, USA}
\affiliation{Kavli Institute for the Physics and Mathematics of the Universe (WPI), The University of Tokyo Institutes for Advanced Study, The University of Tokyo, 5-1-5 Kashiwanoha, Kashiwa, Chiba 277-8583, Japan}
\author{Nao Suzuki}
\affiliation{Kavli Institute for the Physics and Mathematics of the Universe (WPI), The University of Tokyo Institutes for Advanced Study, The University of Tokyo, 5-1-5 Kashiwanoha, Kashiwa, Chiba 277-8583, Japan}
\author{Ichiro Takahashi}
\affiliation{Astronomical Institute, Tohoku University, 6-3 Aramaki Aza-Aoba, Aoba, Sendai, Miyagi 980-8578, Japan}
\author{Masaomi Tanaka}
\affiliation{Astronomical Institute, Tohoku University, 6-3 Aramaki Aza-Aoba, Aoba, Sendai, Miyagi 980-8578, Japan}
\author{Nozomu Tominaga}
\affiliation{Department of Physics, Faculty of Science and Engineering, Konan University, 8-9-1 Okamoto, Kobe, Hyogo 658-8501, Japan}
\affiliation{Kavli Institute for the Physics and Mathematics of the Universe (WPI), The University of Tokyo Institutes for Advanced Study, The University of Tokyo, 5-1-5 Kashiwanoha, Kashiwa, Chiba 277-8583, Japan}
\author{Ken'ichi Nomoto}
\affiliation{Kavli Institute for the Physics and Mathematics of the Universe (WPI), The University of Tokyo Institutes for Advanced Study, The University of Tokyo, 5-1-5 Kashiwanoha, Kashiwa, Chiba 277-8583, Japan}

% HSC external collaborators
\author{Jeff Cooke}
\affiliation{Centre for Astrophysics \& Supercomputing, Swinburne University of Technology, Hawthorn, VIC 3122, Australia}
\author[0000-0002-1296-6887]{Llu\'is Galbany}
\affiliation{Departamento de F\'isica Te\'orica y del Cosmos, Universidad de Granada, E-18071 Granada, Spain}
\author{Santiago Gonz\'alez-Gait\'an}
\affiliation{CENTRA, Instituto Superior T\'ecnico, Universidade de Lisboa, Portugal}
\author{Chien-Hsiu Lee}
\affiliation{National Optical Astronomy Observatory, 950 North Cherry Avenue, Tucson, AZ 85719, USA}
\author{Giuliano Pignata}
\affiliation{Departamento de Ciencias F\'isicas, Universidad Andres Bello, Avda. Rep\'ublica 252, Santiago, 8320000, Chile}
\affiliation{Millennium Institute of Astrophysics (MAS), Nuncio Monse\~nor S\'otero Sanz 100, Providencia, Santiago, Chile}



\begin{abstract}
Some supernovae such as pair-instability supernovae are predicted to have the duration of more than a year in the observer frame. To constrain the rates of supernovae lasting for more than a year, we conducted a long-term deep transient survey using Hyper Suprime-Cam (HSC) on the 8.2m Subaru telescope. HSC is a wide-field (a $1.75~\mathrm{deg^2}$ field-of-view) camera and it can efficiently conduct transient surveys. We observed the same $1.75~\mathrm{deg^2}$ field repeatedly using the \textit{g}, \textit{r}, \textit{i}, and \textit{z} band filters with the typical depth of 26~mag for 4 seasons (from late 2016 to early 2020). Using these data, we searched for transients lasting for more than a year. Two supernovae were detected in 2 continuous seasons, one supernova was detected in 3 continuous seasons, but no transients lasted for all 4 seasons searched. The discovery rate of supernovae lasting for more than a year with the typical limiting magnitudes of 26~mag is constrained to be $1.4^{+1.3}_{-0.7}(\mathrm{stat.}){}^{+0.2}_{-0.3}(\mathrm{sys.})~\mathrm{events~deg^{-2}~yr^{-1}}$. All the long-lasting supernovae we found are likely Type~IIn supernovae and our results indicate that about 40\% of Type~IIn supernovae have long-lasting light curves. No plausible pair-instability supernova candidates lasting for more than a year are discovered. By comparing the survey results and survey simulations, we constrain the luminous pair-instability supernova rate up to $z\simeq 3$ should be of the order of $100~\mathrm{Gpc^{-3}~yr^{-1}}$ at most, which is $0.01-0.1$ per cent of the core-collapse supernova rate.
\end{abstract}



\section{Introduction}\label{sec:introduction}
The past decade saw the dawn of large-scale time-domain astronomy. Many transient surveys, such as Palomar Transient Factory \citep[PTF,][]{law2009ptf}, Panoramic Survey Telescope and Rapid Response System \citep[Pan-STARRS,][]{kaiser2010panstarrs}, Dark Energy Survey \citep[DES,][]{flaugher2015decam}, Asteroid Terrestrial-impact Last Alert System \citep[ATLAS,][]{tonry2018atlas}, and Zwicky Transient Facility \citep[ZTF,][]{bellm2019ztf}, have made astronomers realize that the Universe is far more dynamic in time than previously assumed. Many transient surveys in the past decade pushed the frontier of short-timescale transients to search for, e.g., shock breakout of supernovae (SNe, e.g., \citealt{forster2016hits,forster2018hits}). They discovered many short-timescale transients that were not known a decade ago \citep[e.g.,][]{drout2014fast,tanaka2016hscrapid,arcavi2016fast,pursiainen2018rapid,rest2018fast,prentice2018cow,margutti2019fbot,tampo2020hscrapid,ho2020fast} and provided a constraint on the rates of the short-timescale transient phenomena \citep[e.g.,][]{berger2013}. Even a transient survey with the cadence of 0.5~seconds has been recently performed \citep{richmond2020}.

However, the frontier of the time-domain astronomy is not limited to the short-timescale phenomena. Indeed, many transients are observed to have the duration of years. Some transients have intrinsically long timescale. For example, Type~IIn SN (SN~IIn) 2008iy \citep{miller2010sn2008iy} and HSC16aayt \citep{moriya2019hsc16aayt} had rise times of more than 100~days and their luminosity decline rates were similarly slow. Another SN~IIn 2003ma had a quick ($\simeq 15~\mathrm{days}$) rise but it kept its luminosity for about 1000~days after the rise \citep{rest2011sn2003ma}. iPTF14hls had the spectra similar to those of the standard SNe~IIP but its plateau phase lasted for more than 600~days \citep{arcavi2017iptf14hls,sollerman2019iptf14hls}. 

A transient timescale becomes even longer if it appears at high redshifts thanks to the time dilation. Superluminous SNe (SLSNe, see \citealt{moriya2018slsnreview,gal-yam2019slsnrev} for recent reviews), for example, can be observed up to $z\simeq 5$ with deep optical transient surveys \citep{tanaka2012surveypredic,tanaka2013surveypredic} and last for several years in the observer frame \citep{cooke2008,cooke2012highz,villar2018slsnlsst,nicholl2020slsniin}. Another type of interesting SNe that are predicted to last for several years in the observer frame is pair-instability SNe (PISNe). The existence of PISNe was predicted in the 1960s \citep{1967ApJ...148..803R,1967PhRvL..18..379B} but no conclusive PISNe have been discovered (see \citealt{terreran2017pisncand} for a recent candidate). Some PISNe are predicted to have luminous long-lasting light curves (LCs, e.g., \citealt{scannapieco2005,kasen2011pisn,dessart2013pisn,kozyreva2014pisn,whalen2014pisn,gilmer2017}). Thus, deep and wide transient surveys lasting for many years are ideal for searching for the long sought-after PISNe. The discovery of high-redshift PISNe will provide precious information on massive stars in the early Universe.

Unlike the case of the short-timescale transient surveys, long-term patient monitoring of the same field is required to explore the long-timescale transient phenomena. There have been some transient surveys monitoring the same fields for more than a decade. For example, Catalina Real-time Transient Survey (CRTS) has been observing the same field since 2007 \citep{drake2009crts,drake2019crts}. The limiting magnitudes of such long-lasting transient surveys are, however, rather shallow (e.g., $V\lesssim 20~\mathrm{mag}$ for CRTS, \citealt{drake2009crts}) and they are not suitable for searching for long-lasting faint transients such as high-redshift SLSNe and PISNe. In particular, PISNe are expected to appear preferentially at high redshifts because they require low metallicity environments and we require deep transient surveys to discover them.

Hyper Suprime-Cam (HSC, \citealt{furusawa2018,kawanomoto2018,komiiyama2018,miyazaki2018}), which is on the 8.2m Subaru Telescope and has a field-of-view of $1.75~\mathrm{deg^2}$, is currently the best ground-based instrument to obtain deep and wide optical images (see, e.g., a list of etendue in Figure~1 of \citealt{forster2020}). We previously reported our results of the half-year deep and wide transient survey conducted with HSC at the COSMOS field \citep{capak2007cosmos} in 2016 - 2017 \citep{yasuda2019cosmos}. The half-year survey led to the discovery of many high-redshift SNe~Ia at $z\gtrsim 1$ \citep{yasuda2019cosmos} as well as high-redshift SLSNe \citep{moriya2019shizuca,curtin2019shizuca}. Even after the completion of the half-year survey, we have been monitoring the $1.75~\mathrm{deg^2}$ COSMOS UltraDeep (UD) field for more than 3~years to explore the long-timescale transient phenomena. In this paper, we introduce the long-term time-domain data and present the result of our search for SNe lasting for more than a year. Although we did not obtain spectroscopic information for some long-lasting SNe we discovered, the long duration itself can be used to exclude contamination from normal SNe by searching for targets with durations much longer than normal SNe, allowing us to constrain the PISN and other long-lasting SN rates.

The rest of this paper is organized as follows. We first describe our HSC transient survey data in Section~\ref{sec:description}. We introduce SNe detected for more than a year in Section~\ref{sec:longlastingsne}. We constrain the observational rate of long-lasting SNe in Section~\ref{sec:longlastingsnrate}. Given all the long-lasting SNe we discovered are likely SNe~IIn, we constrain the fraction of long-lasting SNe~IIn among SNe~IIn in Section~\ref{sec:longlastingsniinfraction}. We constrain PISN and SLSN rates based on our survey data in Section~\ref{sec:pisnslsnrates}. We conclude this paper in Section~\ref{sec:conclusions}. We adopt the standard $\Lambda$CDM cosmology with $H_0=70~\mathrm{km~s^{-1}}~\mathrm{Mpc^{-1}}$, $\Omega_M = 0.3$, and $\Omega_\Lambda = 0.7$ throughout this paper.

\begin{figure}
\epsscale{1.2}
\plotone{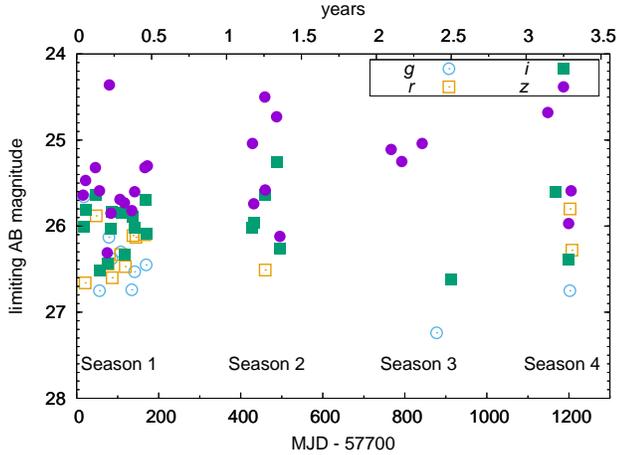}
\caption{
Limiting magnitudes of the HSC long-term transient survey.
No \textit{g} band data were obtained in Season~2 and no \textit{r} band data were obtained in Season~3.
\label{fig:limit}}
\end{figure}

\section{Survey description}\label{sec:description}
The long-term monitoring observations with Subaru/HSC have been conducted in the COSMOS UD field for 4 seasons from late 2016 to early 2020, covering 3.3~years. The COSMOS UD field is centered at (R.A., Dec.) = $(10^h00^m28^s.60$, +02$^\circ$12$'$21.$''$00) and has the area of $1.75~\mathrm{deg^2}$ which is the field-of-view of HSC \citep{aihara2018}. The \textit{g}, \textit{r}, \textit{i}, and \textit{z} band filters are used for the long-term monitoring observations.

Our observational data are listed in Table~\ref{tab:obslog}. Fig.~\ref{fig:limit} summarizes the limiting magnitudes of our transient survey. 
The first season observations (Season 1) were conducted as a part of the Subaru Strategic Program (SSP) with HSC (``Wide-field imaging with Hyper Suprime-Cam: Cosmology and Galaxy Evolution,'' PI: S. Miyazaki, \citealt{aihara2018}). The observations started on 23 November 2016 (UT dates are used in this paper) and ended on 20 June 2017. The overview of this transient survey is presented in \citet{yasuda2019cosmos}. The second season (Season 2) observations were conducted as the intensive program ``HSC Supernova Cosmology Legacy Survey with Hubble Space Telescope'' (S17B-055I, PI: N. Suzuki) from 10 January 2018 to 21 April 2018. No \textit{g} band data were obtained in Season 2. The third season observations (Season 3) were conducted again in the SSP as a back-up target when the sky conditions did not match the SSP criteria. We have the data from 14 December 2018 to 9 May 2019 in Season 3. No \textit{r} band data were obtained in Season 3. The fourth season data (Season 4) were primarily taken during the normal open-use programs ``Variability-based AGN selection with extended COSMOS time-domain survey'' (S20A-073, PI: M. Kokubo) and ``Exploring the long-timescale transient frontier with HSC'' (S20A-042, PI: T. Moriya). Some data were again taken in the SSP as a back-up target. The data in Season 4 were taken from 26 October 2019 to 28 February 2020.

The reference images used to search for transients were taken in 2015 during the SSP as shown in Table~\ref{tab:obslog}. Because the \textit{r} and \textit{i} band filters on HSC were updated in 2016, the reference images for the \textit{r} and \textit{i} band filters were taken using the different filters from those used during the transient survey. However, the difference in the transmission is small\footnote{\url{https://www.subarutelescope.org/Observing/Instruments/HSC/sensitivity.html}} and no significant effects have appeared due to the filter difference.

The data reduction is performed with the same method as described in \citet{yasuda2019cosmos}. In short, the data are reduced by \texttt{hscPipe} \citep{bosch2018}, which is a version of the Vera C. Rubin observatory's Legacy Survey of Space and Time (LSST) stack \citep{ivezic2019,juric2017}. The astrometry and photometry are calibrated relative to the Pan-STARRS1 (PS1) $3\pi$ catalog \citep{schlafly2012,tonry2012,magnier2013,chambers2016}.
The astrometric accuracy is 0.04 arcsec \citep{aihara2018dr1}.
The image subtraction was performed with the method described in \citet{alard1998,alard2000}.

The difference images obtained after the image subtraction were used to identify transient sources. The transient candidates were classified as real or bogus through a machine-learning technique adopting a convolutional neural network (CNN) as described in \citet{yasuda2019cosmos}. If a transient candidate is identified as real in two epochs after the CNN screening, it is regarded as a real transient.

\begin{figure}
 \begin{center}
  \includegraphics[width=0.325\columnwidth]{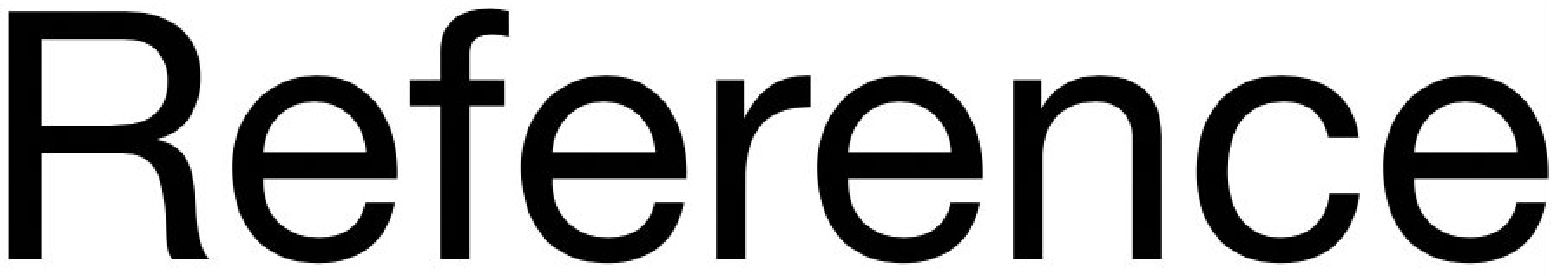}   
  \includegraphics[width=0.325\columnwidth]{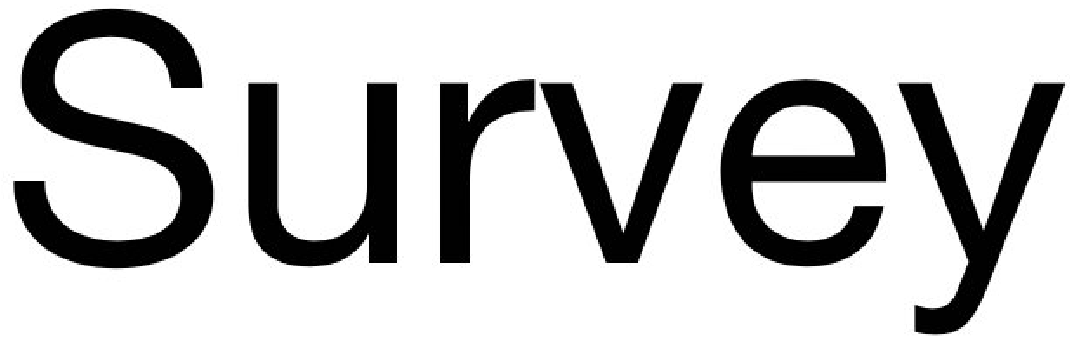}  
  \includegraphics[width=0.325\columnwidth]{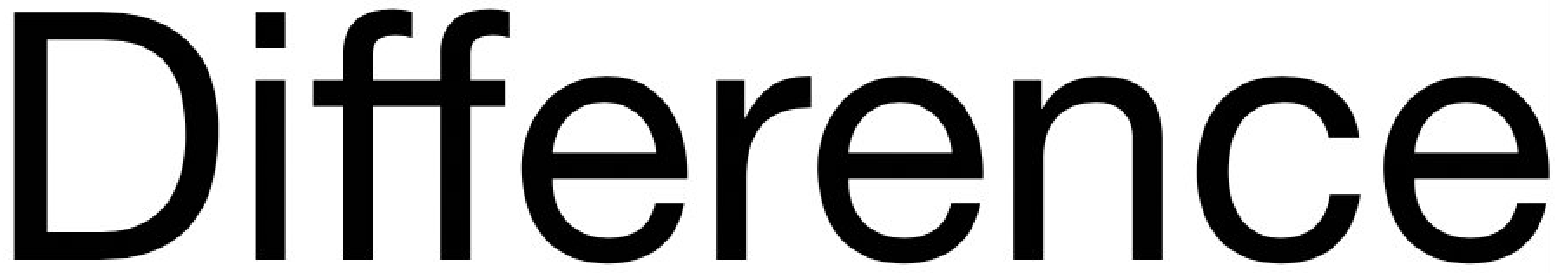}  \\
  \includegraphics[width=0.325\columnwidth]{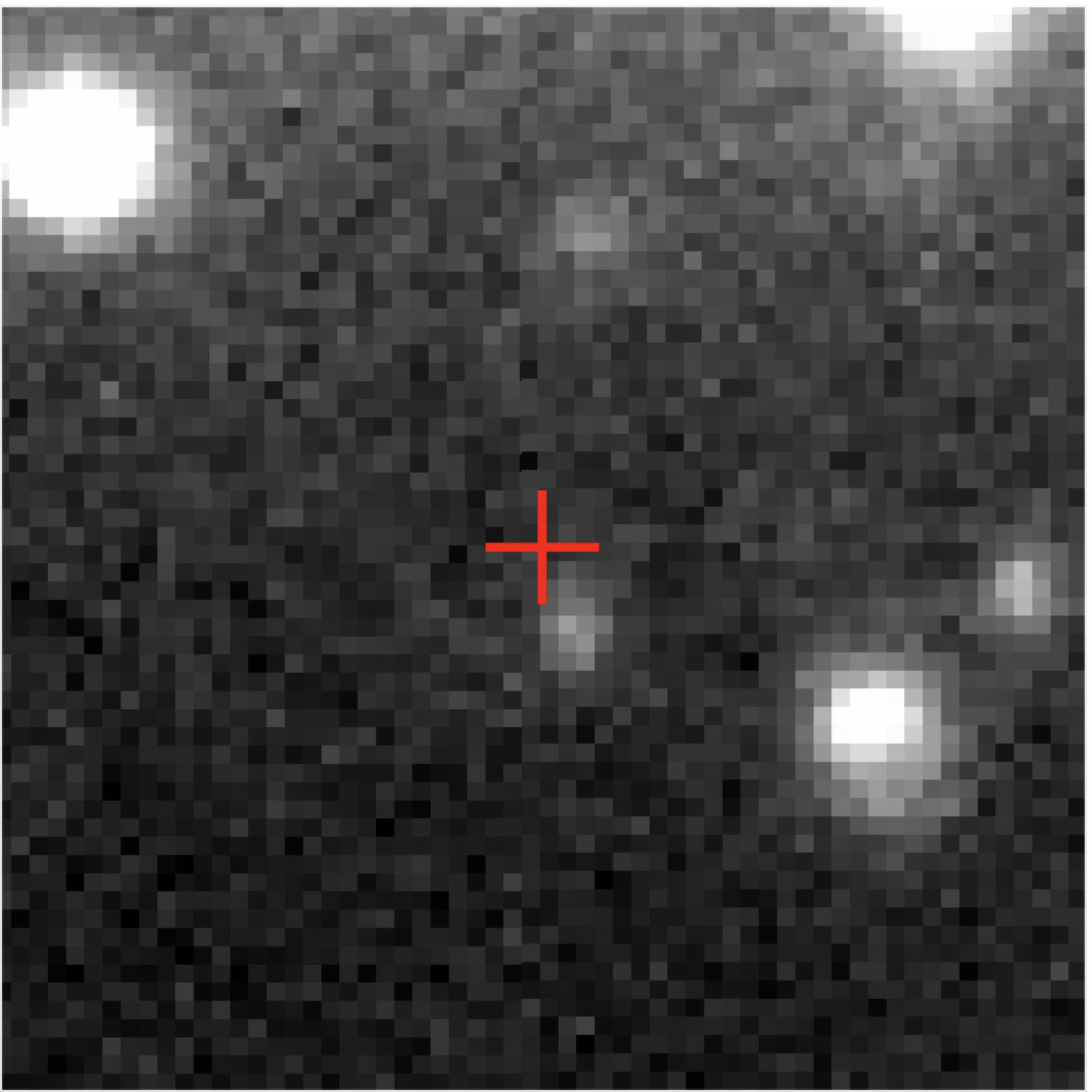}   
  \includegraphics[width=0.325\columnwidth]{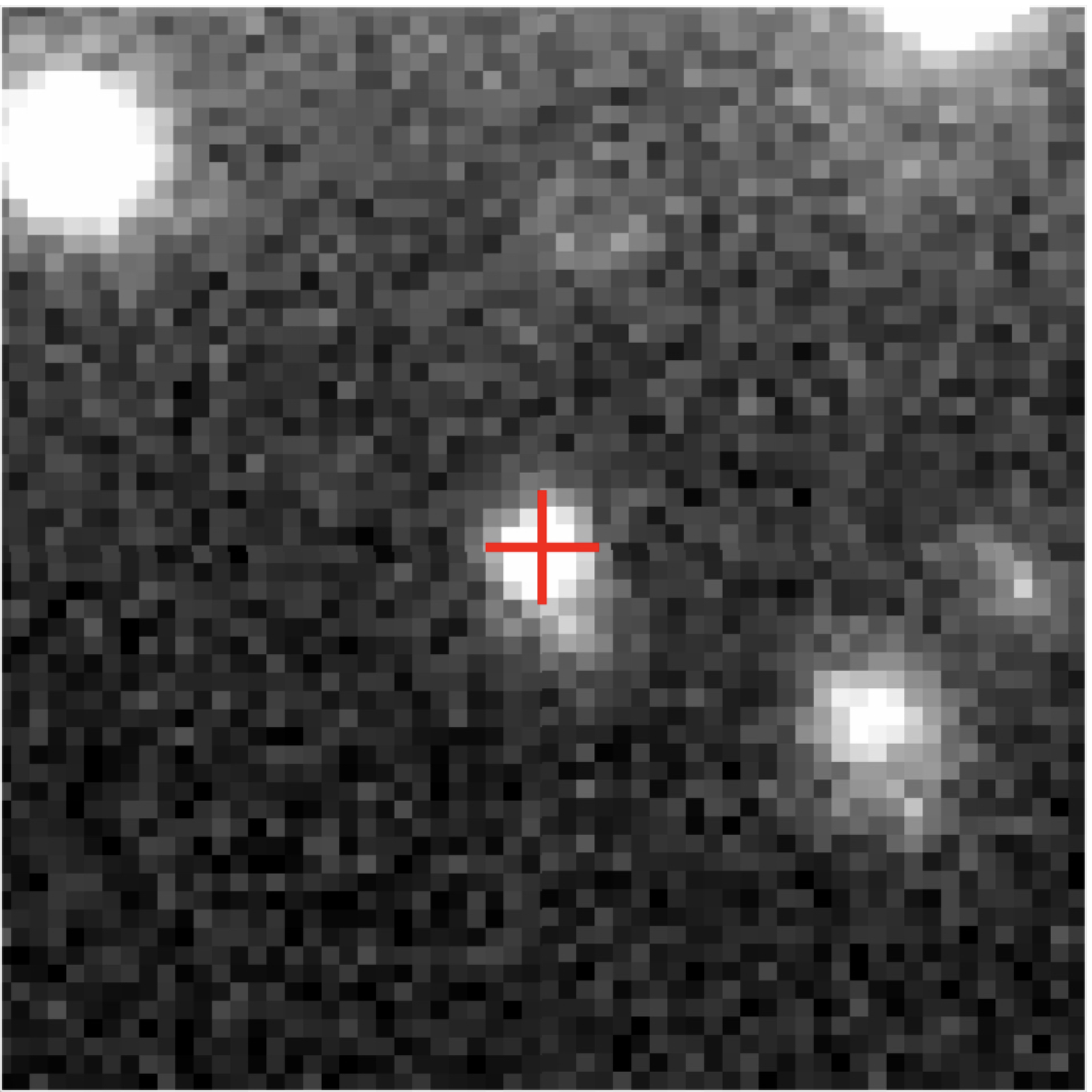}  
  \includegraphics[width=0.325\columnwidth]{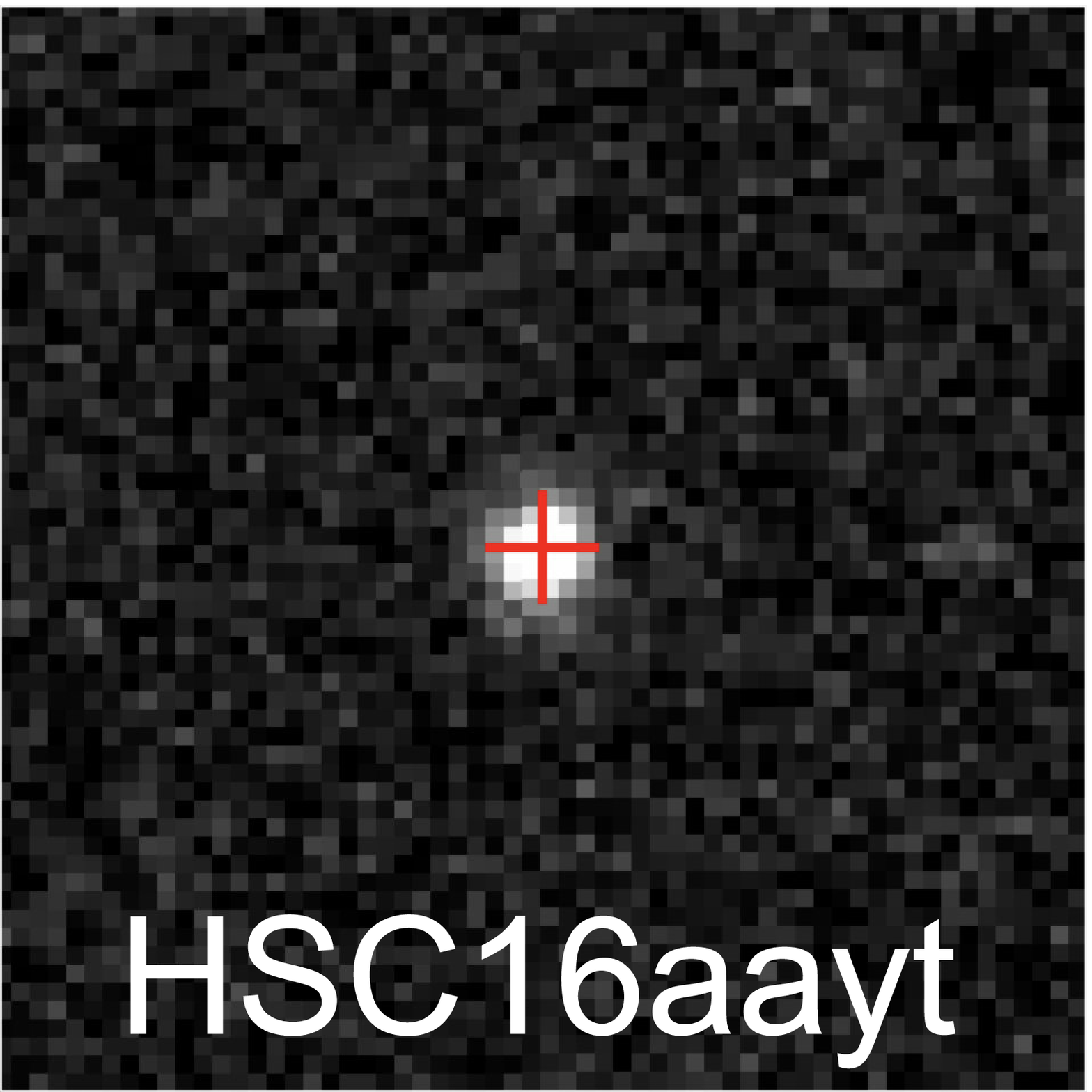}  \\
  \includegraphics[width=0.325\columnwidth]{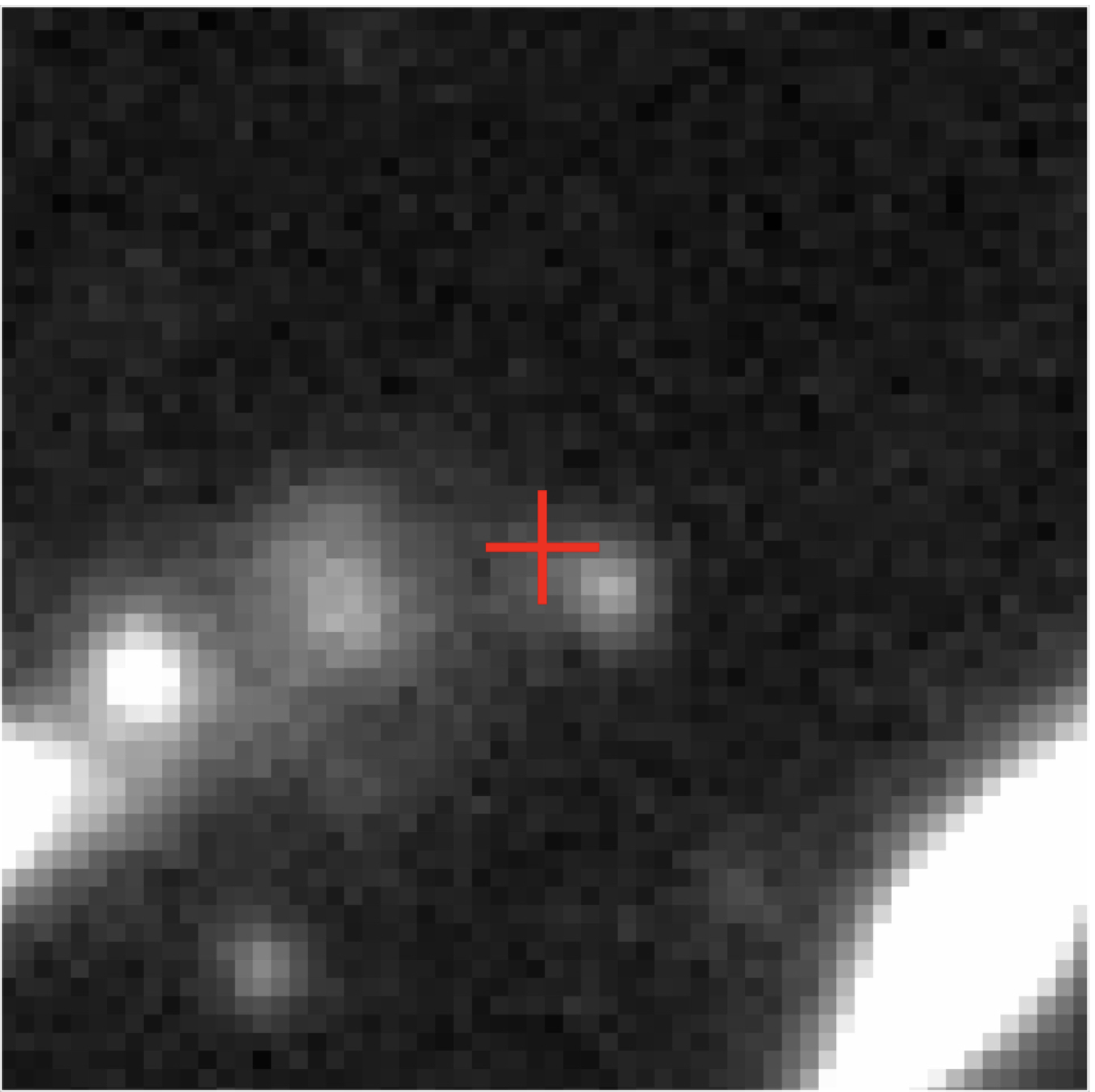}   
  \includegraphics[width=0.325\columnwidth]{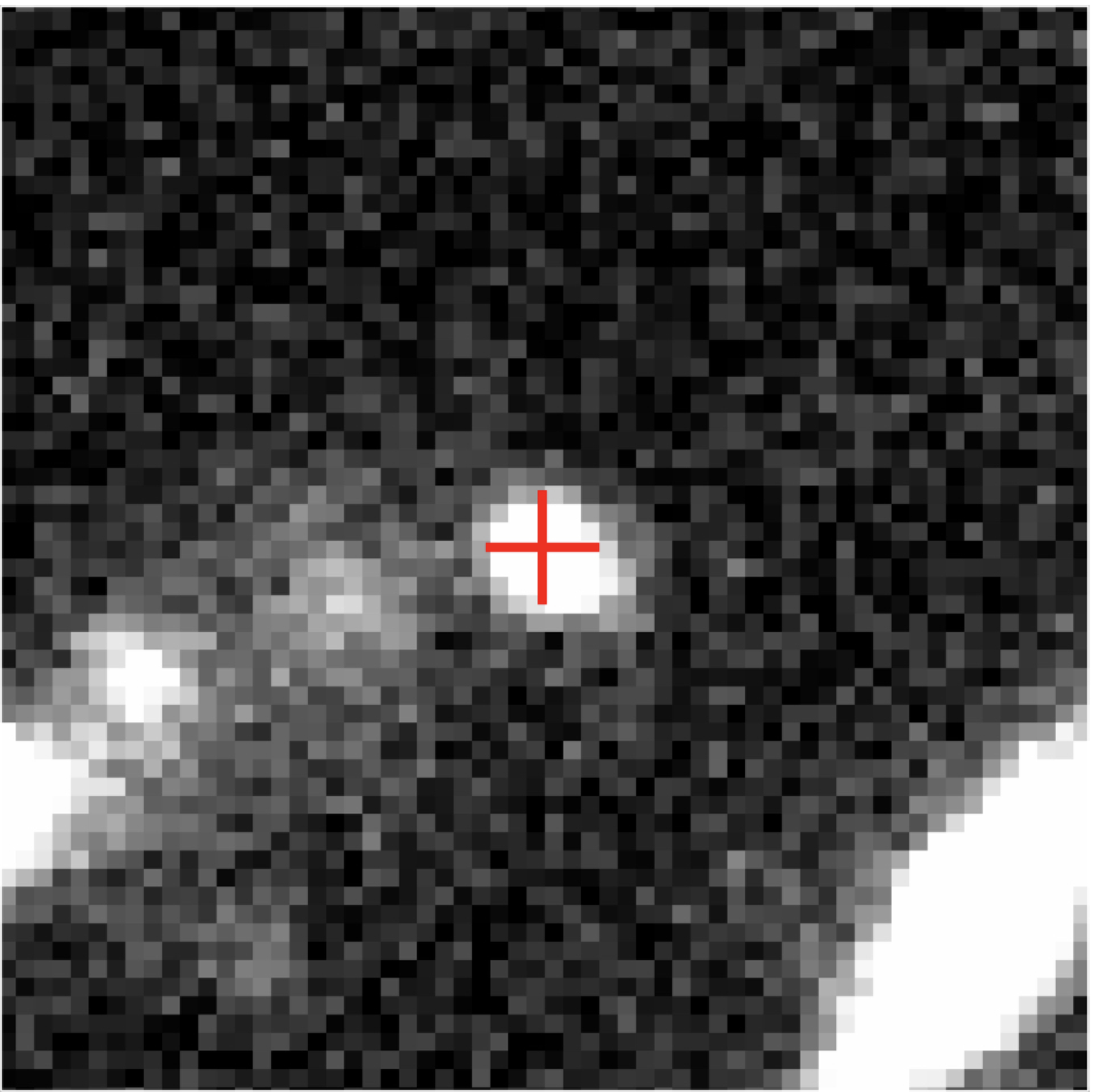}  
  \includegraphics[width=0.325\columnwidth]{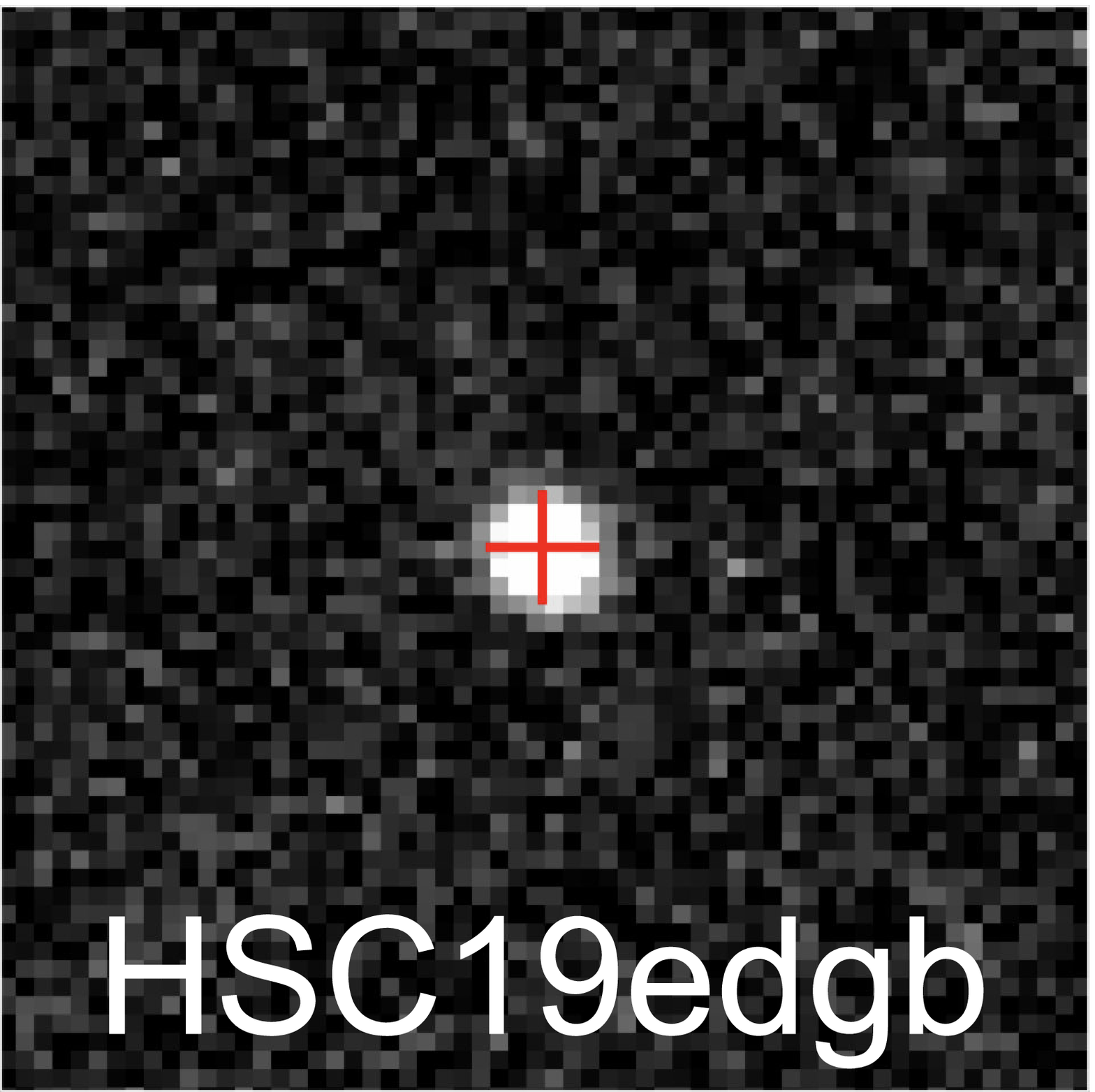}  \\
  \includegraphics[width=0.325\columnwidth]{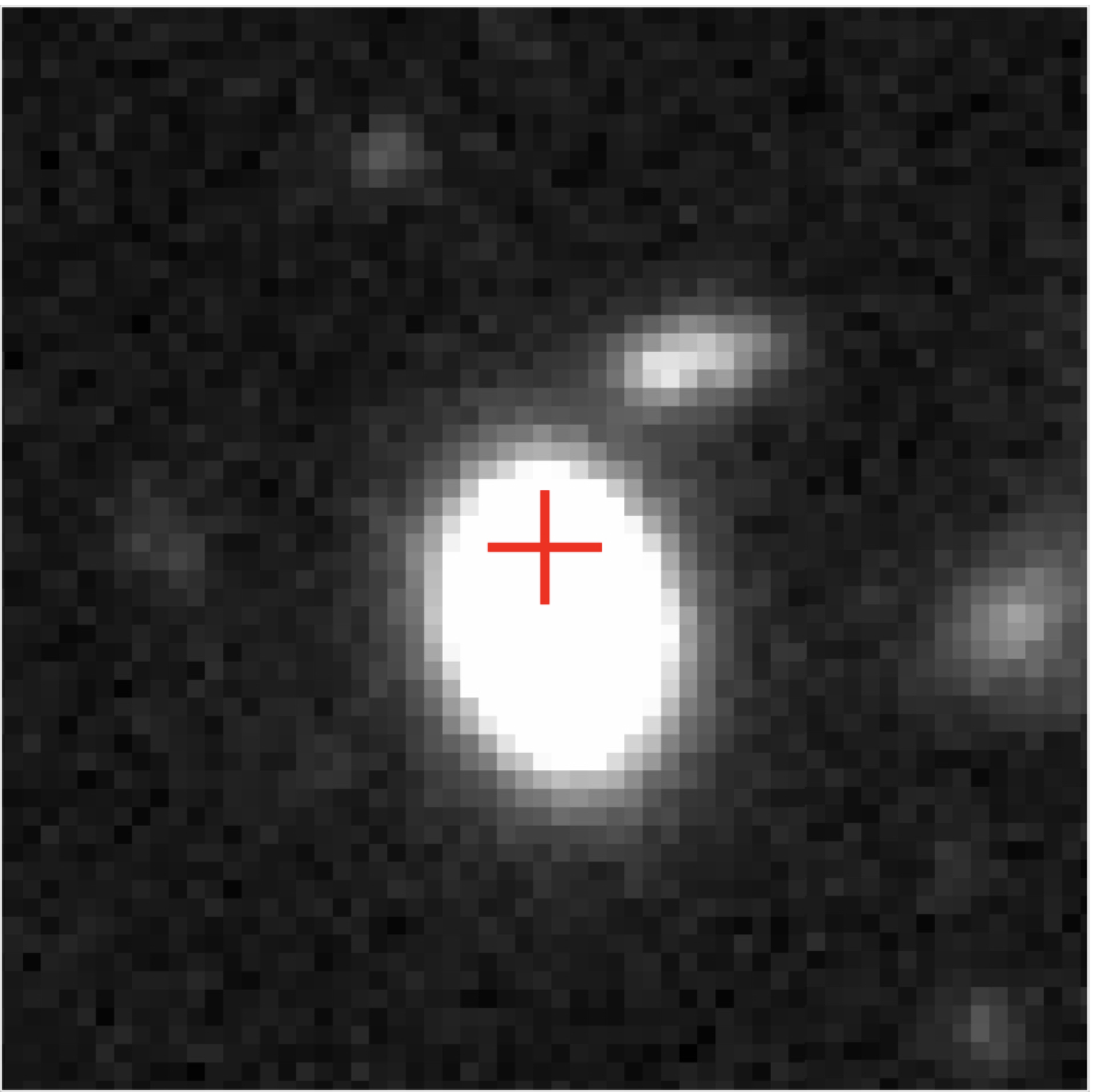}   
  \includegraphics[width=0.325\columnwidth]{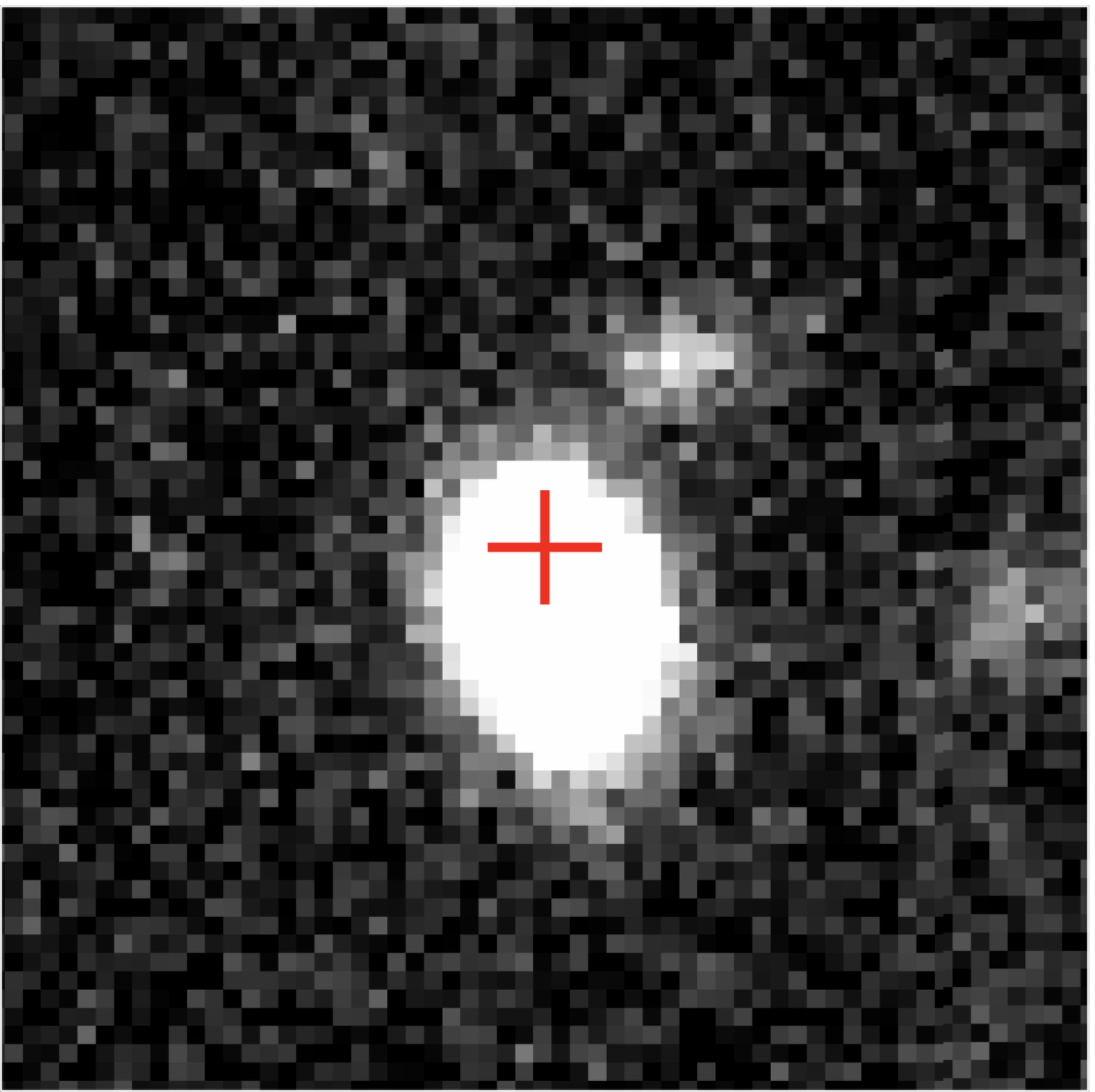}  
  \includegraphics[width=0.325\columnwidth]{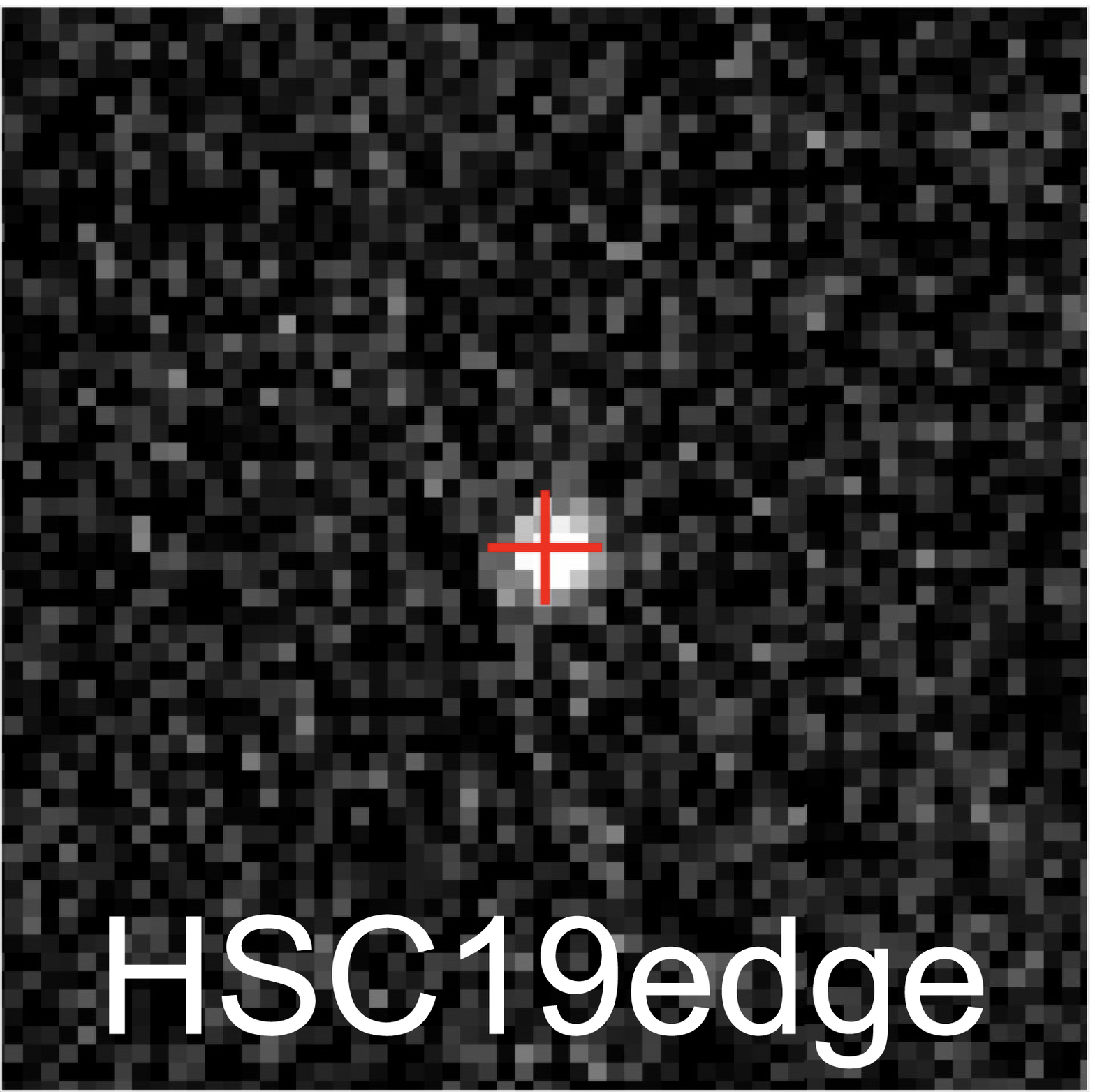}  \\  
 \end{center}
\caption{
The reference images (left), the survey images (middle), and the subtracted images (right) of SNe detected for more than a year during our HSC transient survey.
They are all $z$ band images. The image size is 10$''$ x 10$''$. The SNe are located at the center which is marked with the red cross. North is up and east is left in the images.
}\label{fig:faces}
\end{figure}

To search for SNe lasting for more than a year, we first checked the results of the CNN screening. If a transient is classified as real at any time in one season, it is regarded as a detection in the season. From the transients detected in multiple seasons, we first excluded ``negative'' candidates that have negative flux because they exist in the reference images and we aim at discovering long-lasting SNe that appeared after the reference images were taken. Then, we excluded those identified on top of a point source in the reference image to avoid variable stars. We also exclude those located at the center (within 0.1~arcsec) of their host galaxy to avoid active galactic nucleus (AGN) activities. 
0.1~arcsec corresponds to 0.6~kpc at $z=0.5$ and 0.8~kpc at $z=1-3$. Most SLSNe and SNe~IIn discovered in the local transient surveys have more off-sets from the host galaxy center \citep[e.g.,][]{schulze2020ptfcc} and this criterion is not likely to miss many long-lasting SNe. Transients within 0.1~arcsec are independently studied to investigate AGN activities and any peculiar long-lasting transients can also be identified in the separate AGN study.
After this screening, 2212 long-lasting SN candidates remained. The remaining candidates were mostly bogus caused by the failure of the image subtraction but they were not excluded by the initial CNN screening. We visually checked all the candidates and identified three SNe that are detected for more than a year. They are listed in Table~\ref{tab:listoflongsn} and their images are presented in Fig.~\ref{fig:faces}. We introduce them in the next section.

\begin{deluxetable*}{llcccc}
%\tablenum{1}
\tablecaption{
List of long-lasting SNe.
\label{tab:listoflongsn}}
\tablewidth{0pt}
\tablehead{
\colhead{HSC Name} & \colhead{IAU Name} & \colhead{R.A.} & \colhead{Dec.} &
 \colhead{Redshift} & \colhead{Note}
}
%\decimalcolnumbers
\startdata
\multicolumn6c{\textbf{3 years (4 seasons)}} \\
\multicolumn6c{\textit{none}} \\
\hline
\multicolumn6c{\textbf{2 years (3 seasons)}} \\
HSC16aayt & SN~2016jhm & $10^h02^m20^s.12$ & +02$^\circ$48$'$43.$''$3 & $0.6814$\tablenotemark{a}  & SN~IIn \\
\hline
\multicolumn6c{\textbf{1 year (2 seasons)}} \\
HSC19edgb & AT~2019aadg & $10^h01^m36^s.13$ & +02$^\circ$42$'$25.$''$4 & $0.226^{+0.06}_{-0.01}$\tablenotemark{b}  & SN~IIn? \\
HSC19edge & AT~2018lto & $10^h01^m31^s.55$ & +02$^\circ$48$'$26.$''$5 & $0.33094$\tablenotemark{c}  & SN~IIn?
\enddata
%\tablenotetext{*}{HSC19edge may actually be a 3 year transient if the LC declines with the rate observed in Year 4.}
\tablenotetext{a}{Spectroscopic redshift from the SN spectra \citep{moriya2019hsc16aayt}.}
\tablenotetext{b}{Photometric redshift of the host galaxy in the COSMOS2015 catalog \citep{laigle2016cosmos2015}. Another photometric redshift solution at $z=2.7$ is derived based solely on the HSC photometry (see Section~\ref{sec:19edgb}).}
\tablenotetext{c}{Spectroscopic redshift of the host galaxy in the COSMOS2015 catalog \citep{laigle2016cosmos2015}.}
\end{deluxetable*}

\section{Long-lasting supernovae}\label{sec:longlastingsne}
We introduce our long-lasting SNe in this section. We discuss the discovery rates of long-lasting SNe based on our discovery. When we present the rest-frame magnitudes of SNe, the simple \textit{K} correction of $2.5\log (1+z)$ is applied.

\subsection{3 year-long SNe}\label{sec:3years}
No SNe lasting for 3 years (4 seasons) were identified in our survey.

\subsection{2 year-long SNe}\label{sec:2years}
One SN (HSC16aayt) is detected for 2 years (3 seasons). 

\subsubsection{HSC16aayt}\label{sec:16aayt}
\begin{figure}
\epsscale{1.2}
\plotone{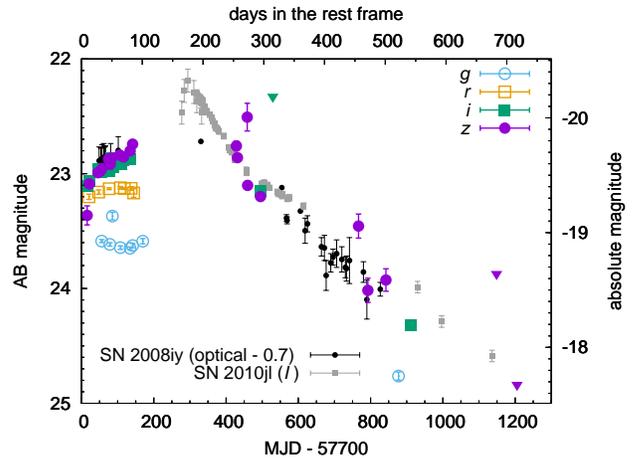}
\caption{
LC of HSC16aayt (SN~IIn) detected for 2 years (3 seasons). Triangles show the limiting magnitudes. 
The optical LC of SN~IIn 2008iy (shifted by $-0.7~\mathrm{mag}$, \citealt{miller2010sn2008iy}) and the \textit{I} band LC of SN~IIn 2010jl \citep{zhang2012sn2010jl} are plotted with the top and right axes for comparison.
Photometric data are available in Table~\ref{tab:hsc16aaty}.
\label{fig:hsc16aayt_lc}}
\end{figure}

HSC16aayt (SN~2016jhm) is discovered at the beginning of our long-term HSC transient survey and identified as a SN~IIn at $z=0.6814$ through spectroscopic follow-up observations \citep{moriya2019hsc16aayt}. The SN is located at $0.71$ arcsec away from the host galaxy center at south west (Fig.~\ref{fig:faces}). The first two season data including spectra have been published in \citet{moriya2019hsc16aayt} and we refer to the paper for the full details of HSC16aayt. It continued to be detected in Season 3 but was below the detection limit in Season 4. The LC of HSC16aayt is shown in Fig.~\ref{fig:hsc16aayt_lc}. The LC decline rate is similar to those of other SNe~IIn (Fig.~\ref{fig:hsc16aayt_lc}).

\subsection{1 year-long SNe}\label{sec:1year}
Two SNe (HSC19edgb and HSC19edge) are detected for 2 seasons and they are 1-year-long SNe in the observer frame. Although we do not have spectroscopic information from the two SNe, their location (offset from the host galaxy center) and LC similarity to those of SNe indicate that they are most likely SNe.

\begin{figure}
\epsscale{1.2}
\plotone{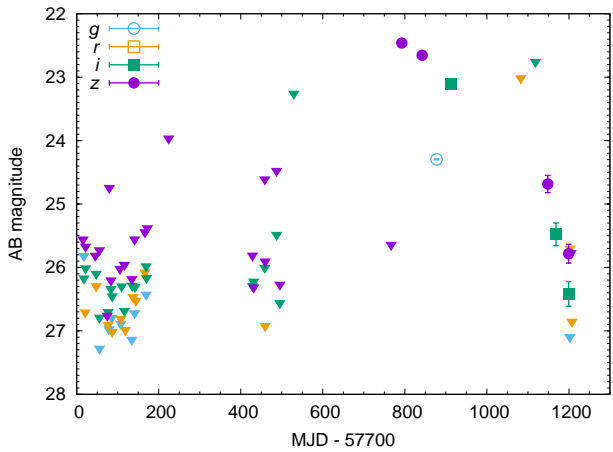}
\caption{
Observed LC of HSC19edgb, which was detected for 1 year (2 seasons). Triangles show the limiting magnitudes. The photometric data are available in Table~\ref{tab:hsc19edgb}.
\label{fig:hsc19edgb_lc}}
\end{figure}

\begin{figure}
\epsscale{2.1}
\plottwo{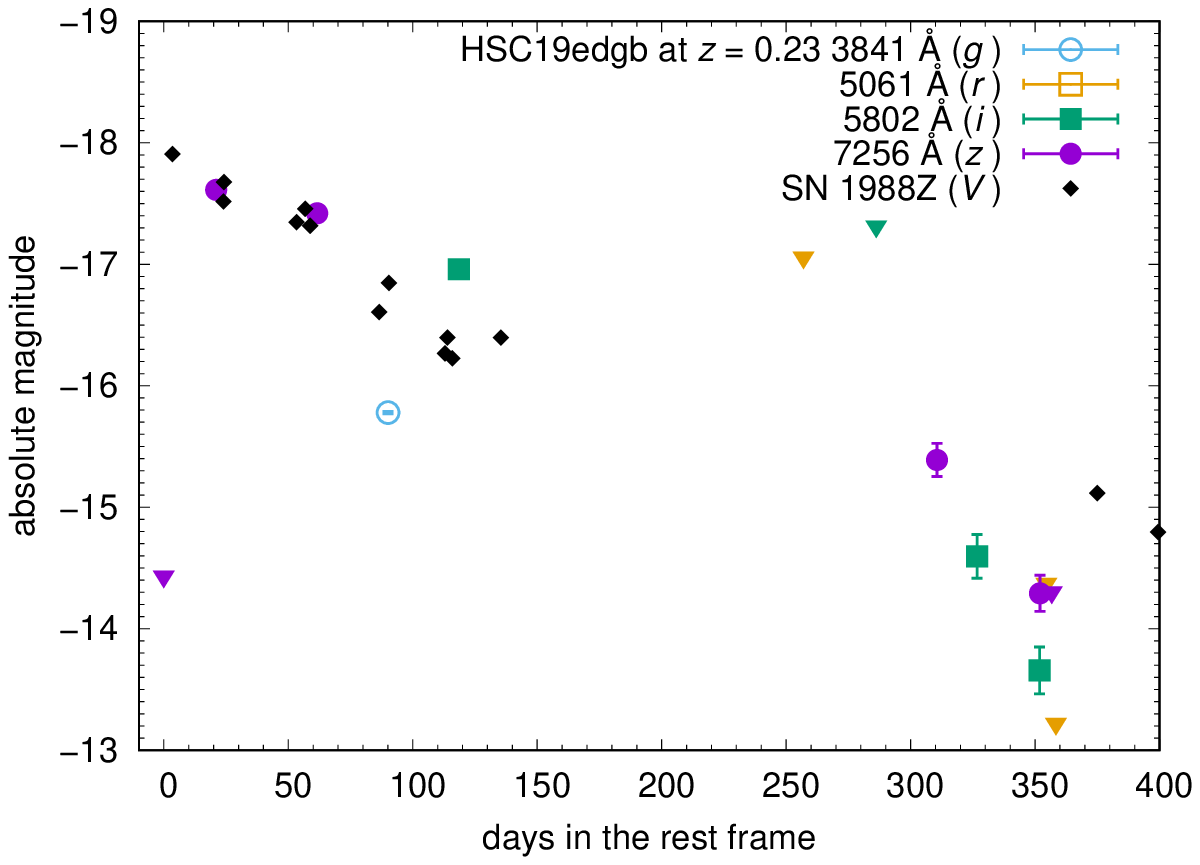}{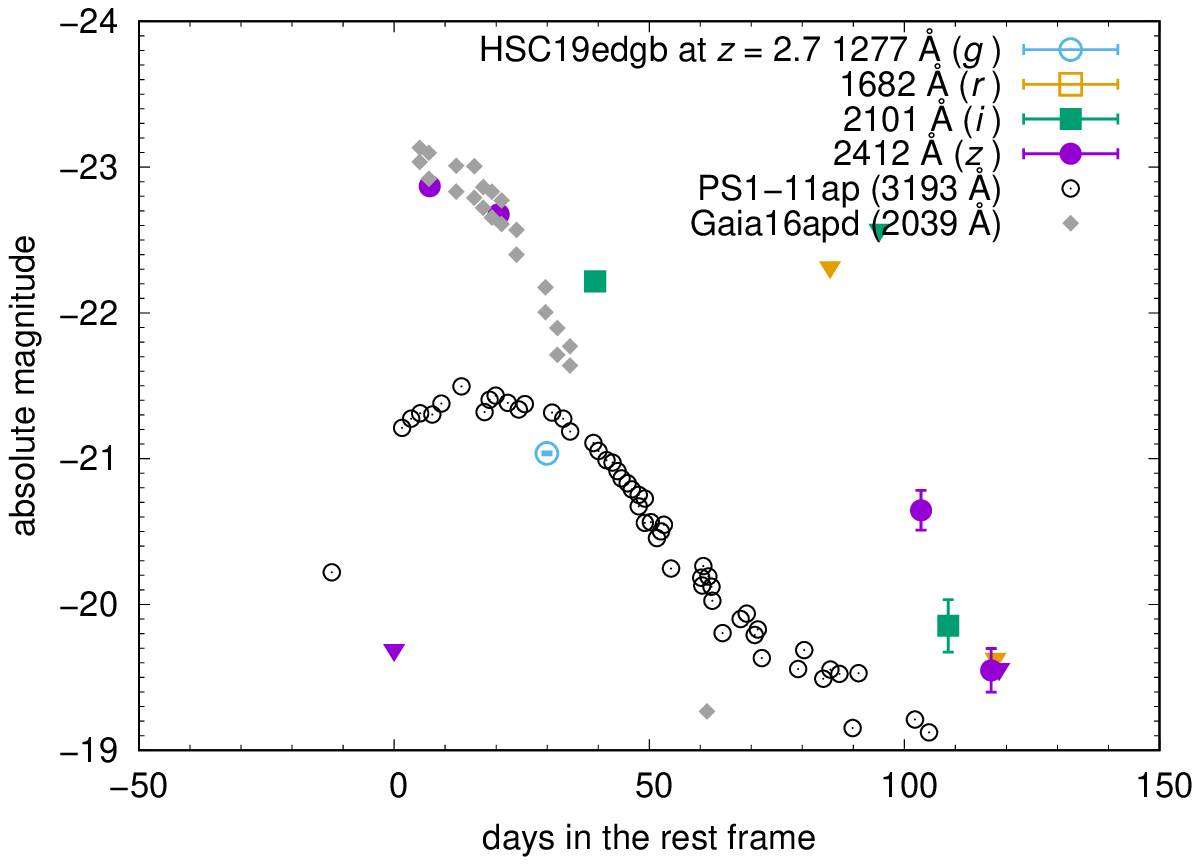}
\caption{
LCs of HSC19edgb in the rest frame in the two possible photometric redshifts at $z=0.23$ and 2.7. The central wavelengths of the HSC filters at each redshift are shown. The time zero is set at the last non-detection at the \textit{z} band before discovery. The LCs of SN~IIn 1988Z \citep{turatto1993sn1988z}, the ultraviolet-bright SLSN Gaia16apd \citep{nicholl2017gaia16apd}, and a SLSN with a typical ultraviolet brightness (PS1-11ap, \citealt{mccrum2014ps1-11ap}) are shown for comparison.
\label{fig:hsc19edgb_lc_rest}}
\end{figure}

\subsubsection{HSC19edgb}\label{sec:19edgb}
HSC19edgb (AT~2019aadg) was first detected on 9 January 2019 in the \textit{z} band. It was not detected in the \textit{z} band on 14 December 2018 when the first data in Season 3 were obtained. Therefore, it was discovered within 26~days after the explosion in the observer frame. The full LC data are presented in Fig.~\ref{fig:hsc19edgb_lc}.

HSC19edgb was located at 0.58~arcsec away towards east from the host galaxy center (Fig.~\ref{fig:faces}). The redshift of the host galaxy is uncertain. The photometric redshift estimated by the COSMOS2015 data is $z = 0.226^{+0.06}_{-0.01}$ \citep{laigle2016cosmos2015}. The photometric redshifts obtained based on the HSC photometry in the reference images are split in two solutions at $z\simeq 0.23$ and 2.7 depending on the method of the photometric redshift estimation \citep{tanaka2018hscphotoz}. Because the COSMOS2015 photometric redshift is based on the multi-wavelength data ranging from radio to X-ray, the lower redshift solution at $z\simeq0.23$ is more reliable. The COSMOS2015 photometric redshift of the extended galaxy at 1.9~arcsec towards east from the transient location is $z = 0.227^{+0.009}_{-0.01}$ and it also supports the low redshift solution.

Fig.~\ref{fig:hsc19edgb_lc_rest} shows the LCs of HSC19edgb at the two photometric redshifts at $z = 0.23$ and 2.7. In the case of $z=0.23$, both luminosity and LC evolution are consistent with those of the slowly-evolving SN~IIn 1988Z \citep{turatto1993sn1988z}.
%Such a slow LC evolution is generally observed in SNe~IIn.
HSC19edgb is naturally explained as a SN~IIn if it is at $z\simeq 0.23$. The peak luminosity is consistent with low-mass PISN models, but their rise times are much longer than that of HSC19edgb \citep{kasen2011pisn}. Thus, it is not likely a PISN at $z\simeq 0.23$.

If HSC19edgb is at $z\simeq 2.7$, it is an ultraviolet-bright transient (Fig.~\ref{fig:hsc19edgb_lc_rest}). Some SLSNe, such as Gaia16apd \citep{nicholl2017gaia16apd,yan2017gaia16apd,kangas2017gaia16apd}, are known to become very bright in ultraviolet and the peak magnitude of HSC19edgb at $z=2.7$ in ultraviolet is consistent with that observed for Gaia16apd. However, the LC decline rate of HSC19edgb in ultraviolet is much slower than that of Gaia16apd. The rise time of HSC19edgb at $z=2.7$ becomes 7~days in the rest frame. Such a fast rise has never been observed in SLSNe, although the ultraviolet LC information in SLSNe is still limited. HSC19edgb is not likely a PISN even at $z\simeq 2.7$ because they are not bright in ultraviolet \citep{kasen2011pisn,dessart2013pisn}.

To summarize, HSC19edgb is more likely a SN~IIn at $z\simeq 0.23$ because (i) the photometric redshift based on broad-frequency information prefers $z\simeq 0.23$ and (ii) the LC evolution is naturally explained as a slowly-evolving SN~IIn at $z\simeq 0.23$. However, the possibility that HSC19edgb is an ultraviolet-bright SLSN at $z\simeq 2.7$ is not excluded. HSC19edgb is not likely a PISN at either $z\simeq 0.23$ or 2.7.

\subsubsection{HSC19edge}\label{sec:19edge}
\begin{figure}
\epsscale{1.2}
\plotone{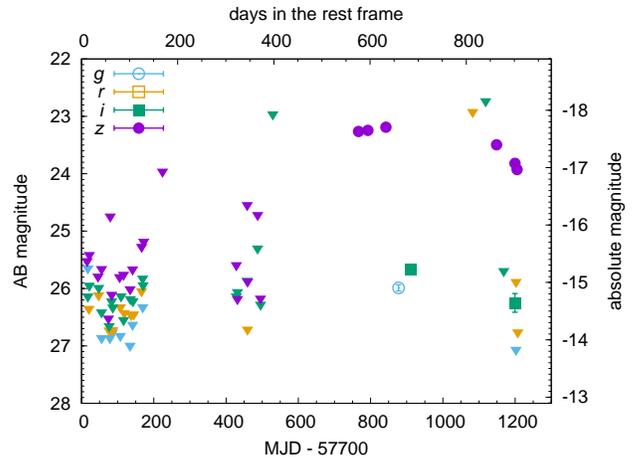}
\caption{
LC of HSC19edge detected for 1 year (2 seasons). Triangles show the limiting magnitudes. The photometry data are provided in Table~\ref{tab:hsc19edge}.
\label{fig:hsc19edge_lc}}
\end{figure}

HSC19edge (AT~2018lto) was detected from the beginning of Season 3. Nothing was detected at the position in Seasons 1 and 2. It was located at 0.67~arcsec away towards north from the host galaxy center (Fig.~\ref{fig:faces}). The spectroscopic redshift of the host galaxy, $z=0.33094$, is available in the COSMSO2015 catalog \citep{laigle2016cosmos2015}. We assign the same redshift to HSC19edge. Fig.~\ref{fig:hsc19edge_lc} shows the LC of HSC19edge.

\begin{figure}
\epsscale{1.2}
\plotone{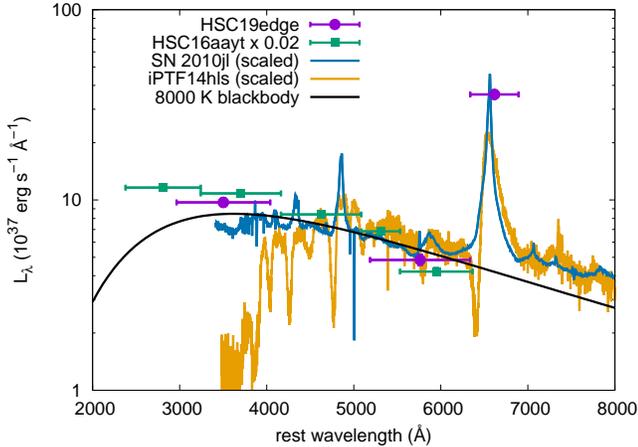}
\caption{
SED of HSC19edge based on the photometry in Season 3. The scaled photometry of SN~IIn HSC16aayt (Section~\ref{sec:16aayt}) 
and the scaled spectra of SN~IIn 2010jl \citep{smith2012sn2010jl} and the peculiar SN~II iPTF14hls \citep{arcavi2017iptf14hls} are shown for comparison. We also show a blackbody SED from the blackbody temperature of 8000~K and radius of $4\times 10^{14}~\mathrm{cm}$ for reference.
\label{fig:hsc19edge_sed}}
\end{figure}

HSC19edge has a flux excess in the \textit{z} band (Fig.~\ref{fig:hsc19edge_lc}). In Fig.~\ref{fig:hsc19edge_sed}, we plot the spectral energy distribution (SED) of HSC19edge based on the \textit{g}, \textit{i}, and \textit{z} band photometry in Season~3. We find that the rest-frame wavelength of the \textit{z} band filter, which is the reddest band we have, matches with the H$\alpha$ wavelength in the rest frame. Therefore, the \textit{z} band flux excess is likely caused by the strong H$\alpha$ emission from HSC19edge. 
A strong H$\alpha$ emission is a generic feature of SNe~II. Among SNe~II, SNe~IIn and some peculiar SNe~II such as iPTF14hls are known to have LCs with a long luminous phase as seen in HSC19edge. Fig.~\ref{fig:hsc19edge_sed} compares the SED of HSC19edge with the spectra of SN~IIn 2010jl and iPTF14hls. The two spectra match well up to around 4000~\AA, but the blue SN~IIn spectrum matches the SED of HSC19edge better at the shorter wavelengths than that of iPTF14hls. Therefore, HSC19edge is more likely to be a SN~IIn, although the possibility of being an iPTF14hls-like SN is not excluded.

\section{Observational rates of long-lasting supernovae}\label{sec:longlastingsnrate}
Based on the long-lasting (more than a year) SN discoveries reported in the previous section, we estimate the observational rate of long-lasting SNe with the survey depth of around 26~mag.

The expected number of events $N$ during our survey can be expressed as
\begin{equation}
    N = \epsilon R t_\mathrm{s} A_\mathrm{s},
\end{equation}
where $\epsilon$ is the discovery efficiency, $R$ is the event rate in the unit of event~$\mathrm{deg^{-2}~yr^{-1}}$, $t_\mathrm{s}$ is the survey duration in the unit of yr, and $A_\mathrm{s}$ is the survey area in the unit of $\mathrm{deg^{2}}$. Our survey conducted with HSC has $A_\mathrm{s}=1.75~\mathrm{deg^{2}}$.

In order to be identified as a multi-season transient during our survey, the transient needs to appear by Season 3 so that we can confirm their long-lasting nature in Season 4. It is not yet clear the transients identified only in Season 4 last for more than one season or not. Therefore, we set $t_s=2.5~\mathrm{yr}$, which is from the beginning of Season 1 to the end of Season 3 (Fig.~\ref{fig:limit}), to estimate the long-lasting SN rate.

The discovery efficiency $\epsilon$ is an uncertain parameter. We performed a mock transient survey simulation to estimate $\epsilon$. We assumed the same observational epochs as in our survey and randomly generated one-year-long SNe from Season~1 to Season~3 ($t_s=2.5~\mathrm{yr}$) that are brighter than the detection limits. We generated $10^6$ one-year-long SNe in the mock survey simulation and found that 54\%\ of them are observed in multiple seasons. For instance, if a one-year-long SN appears shortly after Season~1, it disappears before Season 3 and it is only detected in Season~2. Such a long-lasting transient is missed in our survey. Discovery efficiency for transients lasting for two or more years are likely larger because our observational gaps are about 0.5 years. In addition to the effects caused by the observational epochs, some transients can be missed during the image subtraction process and the CNN screening \citep{yasuda2019cosmos}. The exclusion of the transients from the galaxy center to avoid contamination by AGN activities can also reduce the efficiency, although its effect is likely not significant (Section~\ref{sec:description}). Overall, we set $\epsilon\simeq 0.5\pm 0.1$ based on our mock survey simulation. For reference, we show the rates within 20\% of the assumed efficiency of $0.5$, which is $0.1$, to present the possible systematic uncertainty.

Under the assumptions discussed so far, the three long-lasting SN discovery in our survey sets the observational rate of SNe lasting for more than a year with a survey having the depth of around 26~mag to be
\begin{equation}
    R = 1.4^{+1.3}_{-0.7}(\mathrm{stat.}){}^{+0.2}_{-0.3}(\mathrm{sys.})~\mathrm{events~deg^{-2}~yr^{-1}}.
%    R = 2.3^{+2.2}_{-1.2}(\mathrm{stat.}){}^{+4.6}_{-0.9}(\mathrm{sys.})~\mathrm{events~deg^{-2}~yr^{-1}}.
\end{equation}
The statistical error corresponds to the 84\% confidence limits assuming the Poisson statistics.

\section{Long-lasting SN~IIn rate}\label{sec:longlastingsniinfraction}
The three long-lasting SNe we discovered are likely all SNe~IIn. The LCs of SNe~IIn are heterogeneous \citep[e.g.,][]{kiewe2012,taddia2013,nyholm2020}. Some evolve very quickly and others last for many years. In this section, we constrain the fraction of long-lasting SNe~IIn among SNe~IIn with our survey data, assuming that the three long-lasting SNe we discovered are all SNe~IIn.

Based on the discovered long-lasting SNe~IIn, the volumetric rate of long-lasting SNe~IIn can be roughly estimated as $\sum_{i} (1+z_i)/\epsilon V t_s$, where $V$ is the survey volume and $z_i$ is the redshift of the individual object. We use $\epsilon\simeq 0.5$ and $t_s=2.5~\mathrm{yr}$ as in the previous section.

The peak optical magnitude of most SNe~IIn are brighter than $-17~\mathrm{mag}$ \citep{nyholm2020,richardson2014lumfunc}. Allowing two magnitudes to identify long-lasting SNe~IIn for more than a year in our survey \citep{nyholm2020}, we assume that our survey is complete for the discovery of long-lasting SNe~IIn up to $z\simeq 0.35$ at which the limiting magnitude of 26~mag roughly corresponds to $-15~\mathrm{mag}$. The corresponding survey volume is $V\simeq 4.6\times 10^{-4}~\mathrm{Gpc^{3}}$ with the field-of-view of $1.75~\mathrm{deg^2}$. Because we limit at $z<0.35$, we exclude HSC16aayt at $z=0.68$ in this analysis.

Given all the assumptions discussed so far, we obtain the long-lasting SN rate of $\sim 3000~\mathrm{SNe~Gpc^{-3}~\mathrm{yr^{-1}}}$. We note that we do not take the effect of dust extinction into account in our estimate. SNe~IIn do not tend to appear at the center of the host galaxies \citep{habergham2014,schulze2020ptfcc} and our exclusion of transients from the host galaxy centers would not affect the rate estimate significantly. The total SN~IIn rate in our search volume is $\sim 7000~\mathrm{SNe~Gpc^{-3}~\mathrm{yr^{-1}}}$ in which we assume that the total core-collapse SN rate is $10^5~\mathrm{SNe~Gpc^{-3}~\mathrm{yr^{-1}}}$ \citep{li2011lickrate,dahlen2012ccrateatz1} and the fraction of SNe~IIn in core-collapse SNe is 7\% \citep{shivvers2017ccfrac}. Thus, roughly 40\% of SNe~IIn are estimated to be long-lasting SNe~IIn if we take $\epsilon\simeq 0.5$. About a half of SNe~IIn are found to last long in nearby transient surveys \citep{nyholm2020} and our estimate with $\epsilon\simeq 0.5$ is not far from the local fraction.

\begin{deluxetable*}{cl|ccccc}
%\tablenum{1}
\tablecaption{
Expected numbers of SN discovery from the survey simulations with the SN rate of $100~\mathrm{Gpc^{-3}~yr^{-1}}$.
\label{tab:sumulationresults}
}\tablewidth{0pt}
\tablehead{
%\multicolumn{2}{c|}{SN Rate} & \multicolumn{5}{c|}{$100~\mathrm{Gpc^{-3}~yr^{-1}}$}&\multicolumn{5}{c}{$10~\mathrm{Gpc^{-3}~yr^{-1}}$} \\
%\hline\hline
\multicolumn{2}{c|}{Model} & \colhead{Total\tablenotemark{a}} & \colhead{3 years\tablenotemark{b}} & \colhead{2 years\tablenotemark{c}} & \colhead{1 year\tablenotemark{d}} & \multicolumn{1}{c}{0 years\tablenotemark{e}} %&\colhead{Total\tablenotemark{a}} & \colhead{3 years\tablenotemark{b}} & \colhead{2 years\tablenotemark{c}} & \colhead{1 year\tablenotemark{d}} & \colhead{0 years\tablenotemark{e}} 
}
%\decimalcolnumbers
\startdata
PISN & R250  & 8.9& 	0&	1.4&	3.7&	 3.8\\ %&  0.87&	0&	0.14&	0.36&	0.38 \\
     & R225  & 9.2& 	0&	1.1&	5.0&	 3.0\\ %& 0.85&	0&	0.10&	0.44&	0.30\\
     & R200  & 5.7& 	0&	0.095&	2.5&	 3.1\\ %& 0.60&	0&	0.007&	0.26&	0.33\\
     & R175  & 3.7& 	0&	0&	   0.36&	 3.3\\ %& 0.34&	0&	0&	0.045&	0.30\\
     & R150  & 2.3& 	0&	0&	    0.1&	 2.2\\ %& 0.24&	0&	0&	0.008&	0.23\\
     & B250  & 5.1& 	0&	0.083&	2.5&	 2.5\\ %& 0.55&	0&	0.011&	0.28&	0.26\\
     & B200  & 1.8& 	0&	0&	   0.67&	 1.1\\ %& 0.19&	0&	0&	0.076&	0.12\\
     & He130 & 6.7& 	0&	0.024&	2.5&	 4.2\\ %& 0.67&	0&	0.002&	0.25&	0.42\\
     & He100 & 3.2& 	0&	0&	   0.39&	 2.8\\ %& 0.31&	0&	0&	0.035&	0.28\\
     & He80  & 0.26&	0&	0&    0.005&	0.26\\ %& 0.022&	0&	0&	0&	0.022\\
\hline
SLSN & Slow   &8.5&	    0&	0&	    2.5&  	6.0\\ %&	0.85&	0&	0&	0.24&	0.61 \\
     & SlowUV & 11&  	0&	0&	    3.1&	7.6\\ %&	1.1&	0&	0&	0.34&	0.80 \\
     & Fast   &7.3& 	0&	0&	   0.44&	6.8\\ %&	0.7&	0&	0&	0.045&	0.66\\
     & FastUV &9.5&	    0&	0&	   0.44&	9.0   %&	0.95&	0&	0&	0.049&	0.90
\enddata
\tablenotetext{a}{Total number of detections, which is the sum of the following four columns.}
\tablenotetext{b}{Detected for 4 seasons.}
\tablenotetext{c}{Detected for 3 continuous seasons. The number does not include those detected for 4 seasons.}
\tablenotetext{d}{Detected for 2 continuous seasons. The number does not include those detected for 4 or 3 continuous seasons.}
\tablenotetext{e}{Detected only in a single season.}
%\tablecomments{This table ``hides'' the third column in the \latex\ when compiled.
%The Distance is also centered on the decimals.  Note that when using decimal
%alignment you need to include the {\tt\string\decimals} command before
%{\tt\string\startdata} and all of the values in that column have to have a
%space before the next ampersand.}
\end{deluxetable*}

\begin{figure}
 \begin{center}
  \includegraphics[width=\columnwidth]{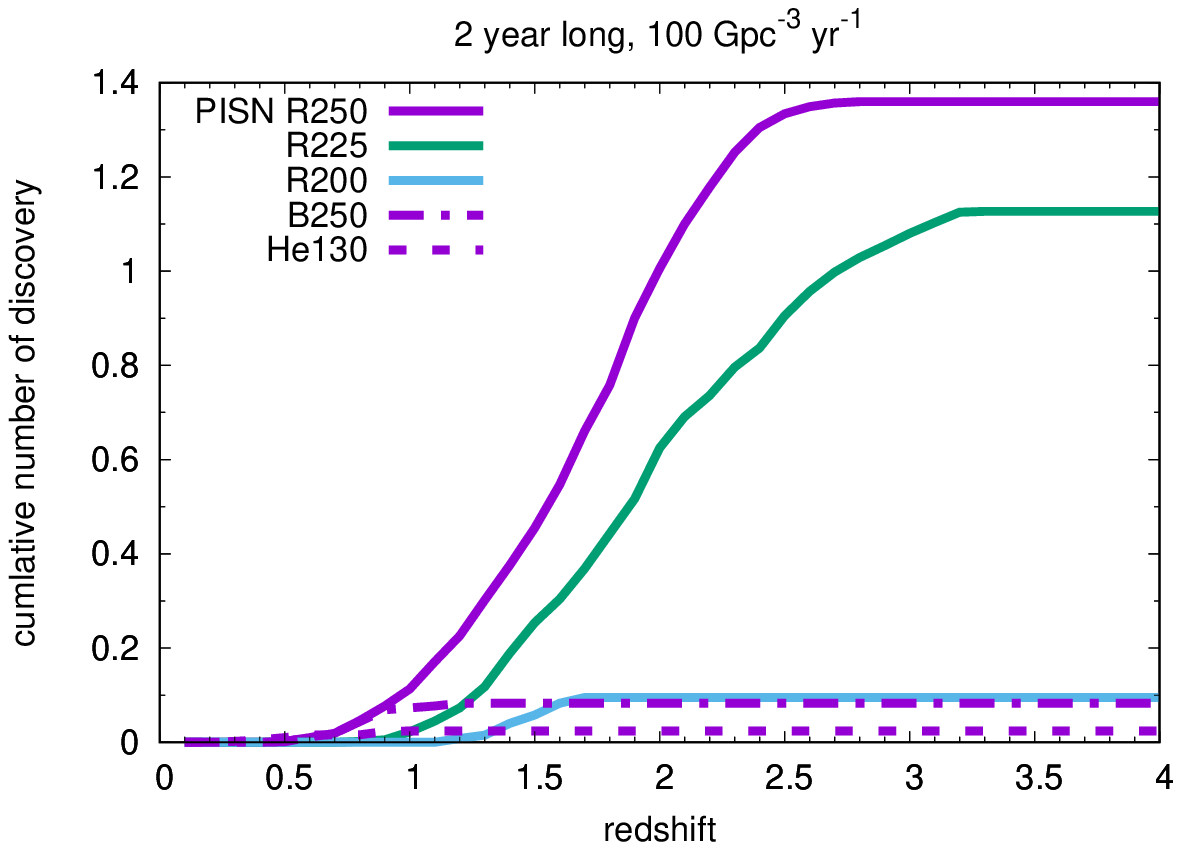}   
  \includegraphics[width=\columnwidth]{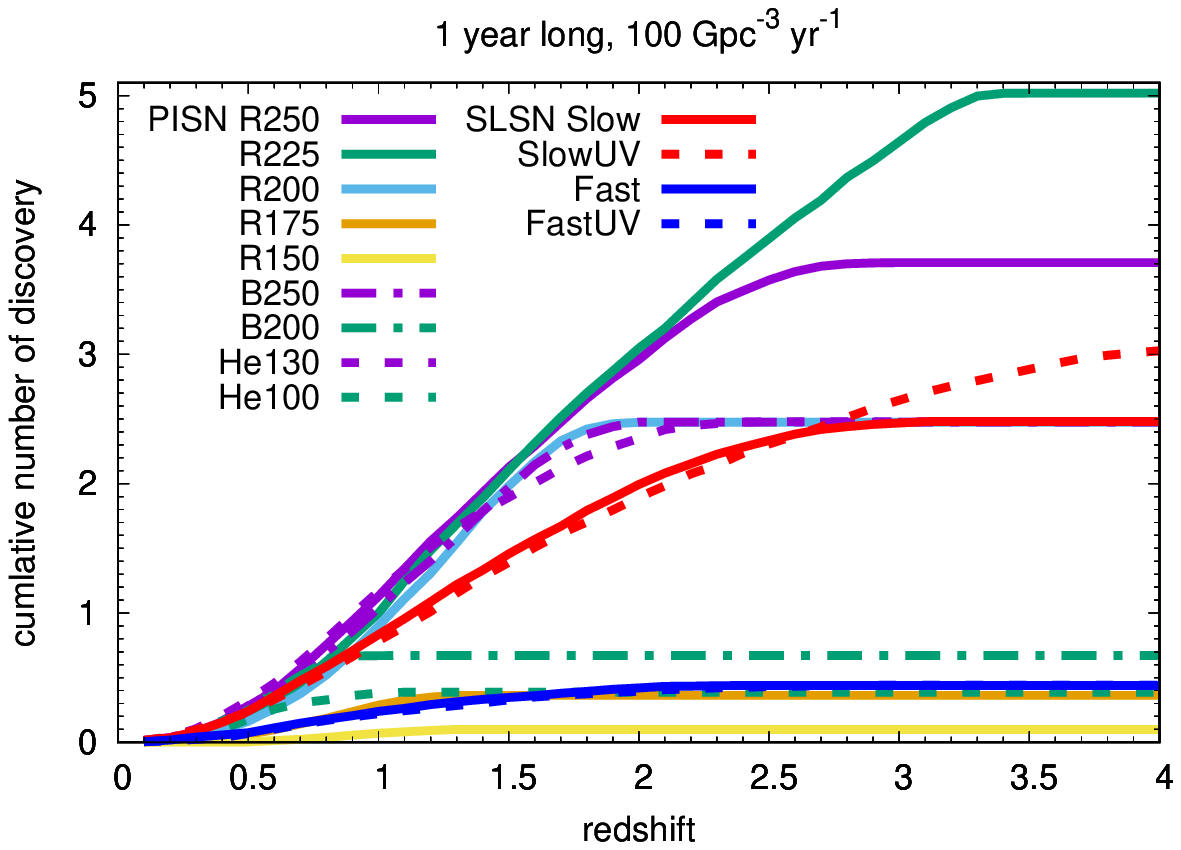}  
  \includegraphics[width=\columnwidth]{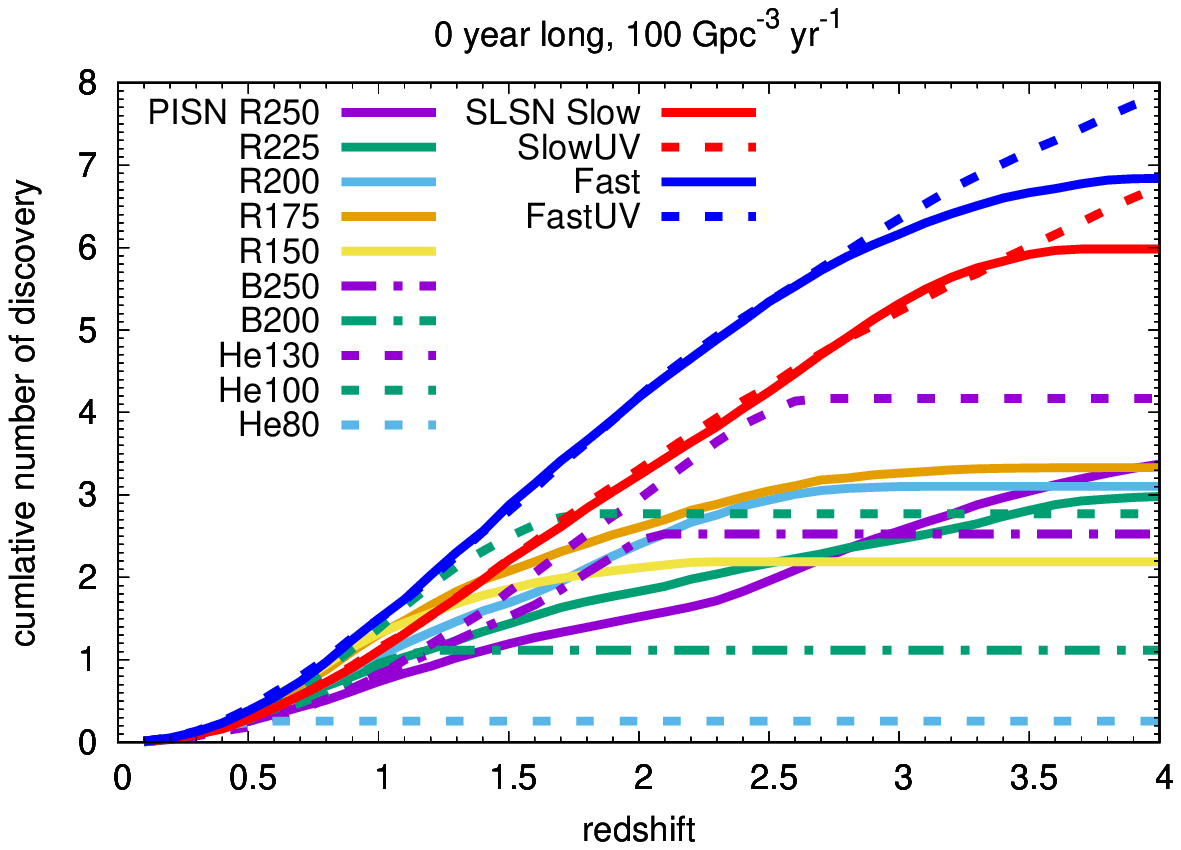}  
 \end{center}
\caption{
Cumulative redshift distribution of the SNe detected in the survey simulations with the rate of $100~\mathrm{Gpc^{-3}~yr^{-1}}$. The top panel is for the 2-year-long SNe, the middle panel is for the 1-year-long SNe, and the bottom panel is for the SNe detected only in one season. The SN models with the total expected number below 0.01 are not plotted.
}\label{fig:sim_redshiftdist}
\end{figure}

\section{Constraints on PISN and SLSN rates}\label{sec:pisnslsnrates}
SLSNe and some PISNe are intrinsically luminous and long-lasting SNe. Thus, they can be observed at high redshifts and their duration can be even longer because of the time dilation. In this section, we conduct the transient survey simulations assuming the same conditions as our HSC transient survey and constrain the PISN and SLSN rates by comparing the survey simulation results and our actual survey results.

Our survey simulations are conducted in the following way. We first make the redshift bin with the interval of 0.1 from $z=0$ to 6. At a given time, we judge if a SN occurs or not based on the assumed SN rate at each redshift bin using the SN rate and the volume in the redshift bin. If a SN occurs, we take the SN LC model from the SN redshift in the observer frame and check if the SN satisfies our detection criteria. If it does, the SN is marked as discovered. We adopt the same limiting magnitudes we had in the real survey presented in Fig.~\ref{fig:limit} and summarized in Table~\ref{tab:obslog}. The survey simulations start long before the observation epochs in order not to miss any SNe in the simulations. If the simulated SN is above the limiting magnitude when we take the reference images, it is excluded because we only searched for long-lasting SNe that became bright after the reference images were taken. The transient survey simulations with the same conditions are performed 1000 times. In each simulation, we count the number of the SN discoveries. We present the average discovery numbers of the 1000 simulations below. The average number is compared with our actual observational results to see the adopted SN model and SN rate are consistent with the observations.

We use PISN LC models calculated by \citet{kasen2011pisn} in our survey simulations. Based on the SED time evolution obtained by \citet{kasen2011pisn}, we calculate the LCs of high-redshift PISNe at the observer frame. We use the R250, R225, R200, R175, R150, B250, B200, He130, He100, and He80 models. This model set includes both faint and luminous PISNe. The number in the model names shows the PISN progenitor mass in the unit of $\mathrm{M_\odot}$. The ``R'' models are red supergiant (RSG) PISN models, the ``B'' models are blue supergiant (BSG) PISN models, and the ``He'' models are helium star PISN models. The SLSN templates we adopt are introduced in Appendix~\ref{sec:app_slsntemp}. We consider SLSNe~I \citep{quimby2011slsn} and refer them as SLSNe here.

All the SN models are assumed to have the same event rate in our simulations. The SN rate is also assumed to be the same at all redshifts for simplicity. We conduct the survey simulations for the event rates of $100~\mathrm{Gpc^{-3}~yr^{-1}}$ and $10~\mathrm{Gpc^{-3}~yr^{-1}}$. The local ($z\simeq 0.2$) SLSN rate is around $30~\mathrm{Gpc^{-3}~yr^{-1}}$ \citep{quimby2013slsnrate} and it increases to around $100~\mathrm{Gpc^{-3}~yr^{-1}}$ at $z\simeq 1$ \citep{prajs2017slsnrate}.
%By comparing the results of the survey simulations to our actual results presented in the previous section, we make a constraint on the PISN and SLSN rates.

The predicted numbers of SN discoveries from our survey simulations with $100~\mathrm{Gpc^{-3}~yr^{-1}}$ are summarized in Table~\ref{tab:sumulationresults}. The discovery numbers with the simulations with $10~\mathrm{Gpc^{-3}~yr^{-1}}$ are reduced by a factor of 10 and the discovery numbers are proportional to the SN rate. We find that the expected numbers of SNe discovered are mostly proportional to the SN rate. No models are detected for 3 years in our simulations. We only find one SN~IIn, which is not a PISN or SLSN, lasting for 2 years in our survey (Section~\ref{sec:2years}).
On average, one two-year-long discovery for both R250 and R225 models is predicted in our survey simulations. Given the small expected number of the detection, we constrain their rates to be of the order of $100~\mathrm{Gpc^{-3}~\mathrm{yr^{-1}}}$ at most.
%The R250 and R225 models are predicted to be detected for 2 years during our survey if their rate is $100~\mathrm{Gpc^{-3}~\mathrm{yr^{-1}}}$. Thus, their rates are constrained to be less than $100~\mathrm{Gpc^{-3}~\mathrm{yr^{-1}}}$.

We find two 1-year-long SNe during our survey, HSC19edgb and HSC19edge. HSC19edge is likely a SN~IIn or a peculiar SN~II (Section~\ref{sec:19edge}). HSC19edgb is most likely to be a SN~IIn, but we cannot exclude the possibility that it is a SLSN (Section~\ref{sec:19edgb}). They are not likely PISNe as discussed in the previous sections. The survey simulations predict that $3-5$ luminous PISN models would be discovered during our survey if their rates are $100~\mathrm{Gpc^{-3}~yr^{-1}}$. The typical redshifts of PISNe discovered in the simulations are $z\simeq 1-3$ (Fig.~\ref{fig:sim_redshiftdist}). Thus, the rates of the luminous PISNe which come from the massive PISN progenitors at $z\simeq 1-3$ are constrained to be less than $100~\mathrm{Gpc^{-3}~yr^{-1}}$ in this case.

The PISN rate is predicted to be of the order of 0.1 per cent of the core-collapse SN rate at $z\lesssim 3$ based on the estimated amount of low-metallicity massive stars \citep[e.g.,][]{langer2007pisnrate,dubuisson2020podsi}. Given the core-collapse SN rate of $\sim  10^{5-6}~\mathrm{Gpc^{-3}~yr^{-1}}$ at these redshifts \citep{li2011lickrate,dahlen2012ccrateatz1}, the nearby PISN rate is predicted to be $\sim 100-1000~\mathrm{Gpc^{-3}~yr^{-1}}$.
Our survey simulations with $100~\mathrm{Gpc^{-3}~yr^{-1}}$ predict a couple of PISN discoveries. Given the small number statistics, the PISN rate is constrained to be of the order of $100~\mathrm{Gpc^{-3}~yr^{-1}}$ at most. Therefore, we conclude that the theoretical prediction is still consistent with our survey results.
It is interesting to note that we expect a couple of PISNe lasting only for one season during our survey if the PISN rate is $\sim 100~\mathrm{Gpc^{-3}~yr^{-1}}$. The one season data need to be investigated carefully because most SNe, including abundant SNe~Ia, last only for a season. We present our investigation of the 1-season data in our forthcoming paper.

We discovered only one potential SLSN lasting for 1 year (HSC19edgb), although it is more likely a SN~IIn. If we take the Fast SLSN model, the number of discoveries (one at most) is consistent with the local SLSN rate of $100~\mathrm{Gpc^{-3}~yr^{-1}}$ with which we expect 0.4 one-year-long SLSN discoveries (Table~\ref{tab:sumulationresults}). The Slow SLSN model predicts around 3 SLSN discoveries in our survey data, which is higher than the actual number of discoveries but non-detection is still consistent with the results given the small number statistics. The Fast model reproduces the \textit{u} band LC behavior better than the Slow model (Appendix~\ref{sec:app_slsntemp}). Thus, it is possible that the ultraviolet LC evolution may be generally fast in SLSNe. The SLSN discoveries in the first season data reported in \citet{moriya2019shizuca,curtin2019shizuca} are consistent with the single season discovery expectation ($\sim 2$ SLSNe per season).

\section{Conclusions}\label{sec:conclusions}
We present the results of our survey for SNe lasting for more than a year by using long-baseline deep (around 26~mag) and wide ($1.75~\mathrm{deg^2}$) HSC time-domain data obtained for 4 seasons from late 2016 to early 2020 in the \textit{g}, \textit{r}, \textit{i}, and \textit{z} band. We discovered no SNe lasting for 3~years, one SNe lasting for 2~years (HSC16aayt), and two SNe lasting for 1~year (HSC19edgb and HSC19edge). Therefore, the discovery rates of SNe lasting for more than a year in the transient surveys with a typical limiting magnitude of 26~mag are estimated to be
$1.4^{+1.3}_{-0.7}(\mathrm{stat.}){}^{+0.2}_{-0.3}(\mathrm{sys.})~\mathrm{events~deg^{-2}~yr^{-1}}$. The statistical error corresponds to the 84\% confidence limits assuming the Poisson statistics and the systematic uncertainty is from the uncertainty in the discovery efficiency.

The three long-lasting SNe we found are all consistent with being a SN~IIn. Assuming that they are all SNe~IIn, we estimate that about 40\% of SNe~IIn have long-lasting LCs. No plausible PISN candidates lasting for more than a year were discovered. By comparing survey simulations and the survey results, we constrain that the PISN rate up to $z\simeq 3$ is less than $100~\mathrm{Gpc^{-3}~yr^{-1}}$. In other words, the PISN rate is less than $0.01-0.1$ per cent of the core-collapse SN rate at these redshifts.

The exploration of the long-timescale (years or more) transient phenomena requires a patient long-term monitoring of the same field. Our HSC data currently have the baseline of around 1000~days. We discovered a couple of long-lasting SNe but longer monitoring the same field is required to explore the frontier of the long-lasting transients. There likely exist many long-lasting rare transients such as PISNe that require longer persistent monitoring of the same field to discover. Our exploration has just started and it is important to keep the monitoring observations for even longer.

\acknowledgments
We thank the anonymous referee for constructive comments that improved this paper.
T.J.M. is supported by the Grants-in-Aid for Scientific Research of the Japan Society for the Promotion of Science (JP17H02864, JP18K13585, JP20H00174).
J.C. would like to acknowledge funding by the Australian Research Council Centre of Excellence for Gravitational Wave Discovery (OzGrav), CE170100004.
L.G. was funded by the European Union's Horizon 2020 research and innovation programme under the Marie Sk\l{}odowska-Curie grant agreement No. 839090. This work has been partially supported by the Spanish grant PGC2018-095317-B-C21 within the European Funds for Regional Development (FEDER). G.P. acknowledge support from the Ministry of Economy, Development, and Tourism’s Millennium Science Initiative through grant IC120009, awarded to The Millennium Institute of Astrophysics, MAS.

This work is supported by the Japan Society for the Promotion of Science Open Partnership Bilateral Joint Research Project between Japan and Chile (JPJSBP120209937).

The Hyper Suprime-Cam (HSC) collaboration includes the astronomical communities of Japan and Taiwan, and Princeton University.  The HSC instrumentation and software were developed by the National Astronomical Observatory of Japan (NAOJ), the Kavli Institute for the Physics and Mathematics of the Universe (Kavli IPMU), the University of Tokyo, the High Energy Accelerator Research Organization (KEK), the Academia Sinica Institute for Astronomy and Astrophysics in Taiwan (ASIAA), and Princeton University.  Funding was contributed by the FIRST program from the Japanese Cabinet Office, the Ministry of Education, Culture, Sports, Science and Technology (MEXT), the Japan Society for the Promotion of Science (JSPS), Japan Science and Technology Agency  (JST), the Toray Science  Foundation, NAOJ, Kavli IPMU, KEK, ASIAA, and Princeton University.

This paper makes use of software developed for the Large Synoptic Survey Telescope. We thank the LSST Project for making their code available as free software at \url{http://dm.lsst.org}.

This paper is based on data collected at the Subaru Telescope and retrieved from the HSC data archive system, which is operated by Subaru Telescope and Astronomy Data Center (ADC) at NAOJ. Data analysis was in part carried out with the cooperation of Center for Computational Astrophysics (CfCA), NAOJ.

The Pan-STARRS1 Surveys (PS1) and the PS1 public science archive have been made possible through contributions by the Institute for Astronomy, the University of Hawaii, the Pan-STARRS Project Office, the Max Planck Society and its participating institutes, the Max Planck Institute for Astronomy, Heidelberg, and the Max Planck Institute for Extraterrestrial Physics, Garching, The Johns Hopkins University, Durham University, the University of Edinburgh, the Queen’s University Belfast, the Harvard-Smithsonian Center for Astrophysics, the Las Cumbres Observatory Global Telescope Network Incorporated, the National Central University of Taiwan, the Space Telescope Science Institute, the National Aeronautics and Space Administration under grant No. NNX08AR22G issued through the Planetary Science Division of the NASA Science Mission Directorate, the National Science Foundation grant No. AST-1238877, the University of Maryland, Eotvos Lorand University (ELTE), the Los Alamos National Laboratory, and the Gordon and Betty Moore Foundation.

The authors wish to recognize and acknowledge the very significant cultural role and reverence that the summit of Maunakea has always had within the indigenous Hawaiian community.  We are most fortunate to have the opportunity to conduct observations from this mountain.


\vspace{5mm}
\facilities{Subaru(HSC)}


\software{
%\texttt{MIZUKI} \citep{tanaka2015mizuki},
\texttt{hscPipe} \citep{bosch2018}
          }



\appendix
\section{SLSN light-curve templates}\label{sec:app_slsntemp}
We describe the SLSN LC templates used for the survey simulations. The SLSNe we consider in this paper are Type~I SLSNe that do not have hydrogen emission features \citep{quimby2011slsn}. We take the similar approach to make the SLSN templates as in \citet{prajs2017slsnrate}. We use the magnetar-powered model \citep{kasen2011pisn,woosley2010magnetar} to make the SLSN LC templates. Although the magnetar model is not necessarily the complete model for SLSNe \citep[][for a recent review]{moriya2018slsnreview}, it reproduces the SLSN LCs at around the peak well \citep[e.g.,][]{inserra2013magnetar,wang2015slsnmagnetar,nicholl2017mosfitmagnetar}.

\begin{figure}[ht!]
\epsscale{2.1}
\plottwo{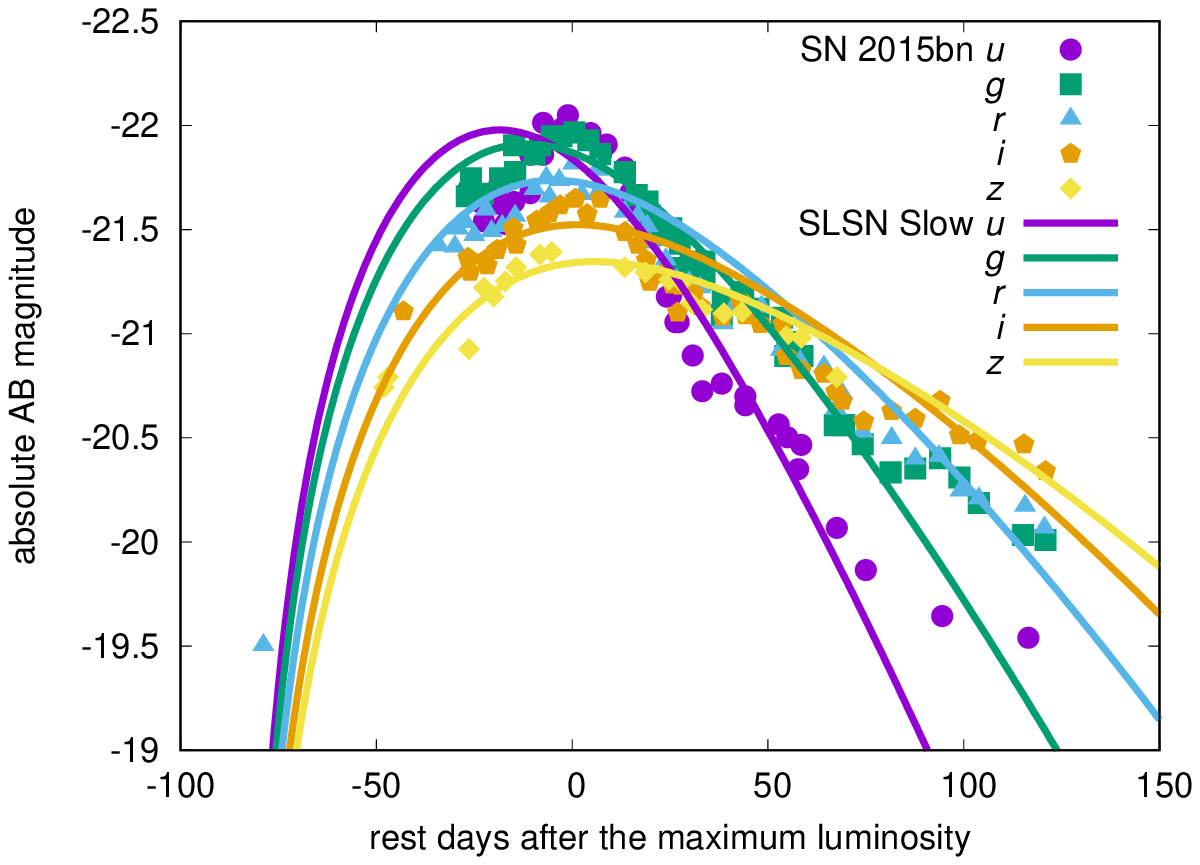}{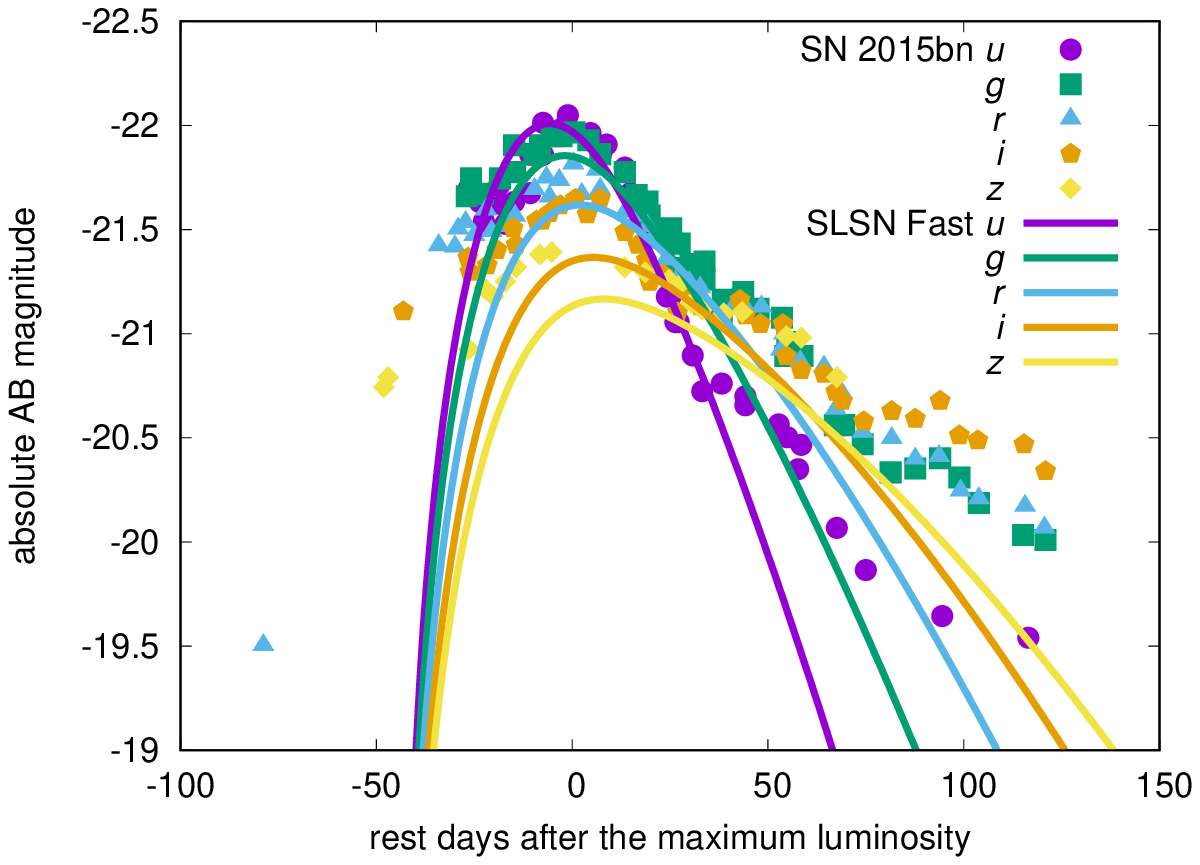}
\caption{
The Slow SLSN LC template (top) and the Fast SLSN LC template (bottom) compared with the observed LCs of SN~2015bn \citep{nicholl2016sn2015bnearly}.
\label{fig:slsntemplates}}
\end{figure}

We take the same approach described in \citet{inserra2013magnetar} to obtain the multi-color LCs of magnetar-powered SNe. The bolometric luminosity is obtained by assuming the central energy input through the dipole radiation,
\begin{equation}
    L_\mathrm{dipole}(t)=\frac{E_p}{t_p}\left(1+\frac{t}{t_p}\right)^{-2},
\end{equation}
where $E_p$ is the initial rotational energy of the magnetar and $t_p$ is the spin-down timescale of the magnetar. Assuming the momentum of inertia of the magnetar is $\simeq 10^{45}~\mathrm{g~cm^2}$, $E_p\simeq 2\times 10^{52} P_\mathrm{ms}^{-2}$, where $P_\mathrm{ms}$ is the initial rotational period scaled with 1~ms. Similarly, assuming the magnetar radius of $\simeq 10~\mathrm{km}$, $t_p\simeq 4.1\times 10^5 B_{14}^{-2}P_\mathrm{ms}^{2}$, where $B_{14}$ is the dipole magnetic field strength of the magnetar. We adopt the semi-analytic model of \citet{arnett1982} (see also \citealt{chatzopoulos2012analc}) to obtain the bolometric LC for which we assume a diffusion time $\tau_d$ in the ejecta. We use our own code to calculate the magnetar bolometric LC \citep{moriya2017magnetar}. Then, we estimate the location of the photosphere assuming the broken power-law ejecta structure ($\rho_\mathrm{ejecta}\propto r^{-10}$ at the outer ejecta and $\rho_\mathrm{ejecta}\propto r^{-1}$ at the inner ejecta) and the ejecta opacity of $0.1~\mathrm{cm^2~g^{-1}}$. Given the bolometric luminosity and photospheric radius, we assume the blackbody function to obtain the SED. The optical SEDs of SLSNe around the peak luminosity match the blackbody function well but the SEDs below $\simeq 3000$~\AA\ are often suppressed \citep[e.g.,][]{vreeswijk2014iptf13ajg,nicholl2017gaia16apd,yan2017gaia16apd}. Therefore, we show the models with and without the ultraviolet suppression. We suppress the blackbody function in the same way as in \citet{prajs2017slsnrate} in which the ultraviolet suppression is estimated based on the SLSN iPTF13ajg \citep{vreeswijk2014iptf13ajg}.

We set $P_\mathrm{ms}=1.5$ and $B_{14}=0.3$ in making the SLSN templates. We use two $\tau_d$: 45~days (the ``Slow'' models) and 25~days (the ``Fast'' models). These values are the typical values obtained when the SLSN LCs are fitted by the magnetar model \citep[e.g.,][]{prajs2017slsnrate,nicholl2017mosfitmagnetar}. The ``Slow'' and ``Fast'' models are presented in Fig.~\ref{fig:slsntemplates}. We can find that the peak magnitudes of our models match well to those of the well-obsereved SLSN SN~2015bn \citep{nicholl2016sn2015bnearly}. The \textit{g}, \textit{r}, \textit{i}, and \textit{z} band LC evolution is well reproduced by the Slow model, while the \textit{u} band LC evolution is well reproduced by the Fast model.

The Slow and Fast models presented so far assume the ultraviolet suppression. We also adopt the models without the ultraviolet suppression in our simulations. They are referred as the ``SlowUV'' and ``FastUV'' models.

\section{Data tables}\label{sec:app_data}
We provide the log of the HSC observations in Table~\ref{tab:obslog}. The photometry data of HSC16aayt, HSC19edgb, and HSC19edge are listed in Tables~\ref{tab:hsc16aaty}, \ref{tab:hsc19edgb}, and \ref{tab:hsc19edge}, respectively.

\begin{longdeluxetable}{llccc}
\tablenum{A.1}
\tablecaption{
Log of the HSC observations.
\label{tab:obslog}}
\tablewidth{0pt}
\tablehead{
\colhead{Date} & \colhead{MJD} & \colhead{Filter} &
 \colhead{Seeing} & \colhead{Limit} \\
 \colhead{} & \colhead{} & \colhead{} &
 \colhead{($''$)} & \colhead{(AB mag)}
}
%\decimalcolnumbers
\startdata
\multicolumn5c{\textbf{Reference}} \\
2015-06-04 & 57177.75 & \textit{g} & 0.81 & \nodata \\
2015-03-06 & 57087.75 & \textit{r} & 0.63 & \nodata \\
2015-01-26 & 57048.44 & \textit{i} & 0.60 & \nodata \\
2015-05-08 & 57150.73 & \textit{z} & 0.59 & \nodata \\
\hline
\multicolumn5c{\textbf{Season 1 (2016 - 2017)}} \\
2016-11-23 & 57715.54 & \textit{z} & 0.71 & 25.64 \\
2016-11-25 & 57717.57 & \textit{g} & 1.09 & 25.66 \\
2016-11-25 & 57717.62 & \textit{i} & 0.80 & 26.01 \\
2016-11-28 & 57720.60 & \textit{r} & 0.76 & 26.66 \\
2016-11-29 & 57721.55 & \textit{i} & 1.15 & 25.81 \\
2016-11-29 & 57721.60 & \textit{z} & 1.04 & 25.47 \\
2016-12-23 & 57745.56 & \textit{z} & 1.05 & 25.32 \\
2016-12-25 & 57747.53 & \textit{r} & 1.12 & 25.88 \\
2016-12-25 & 57747.62 & \textit{i} & 1.23 & 25.64 \\
2017-01-02 & 57755.45 & \textit{z} & 0.73 & 25.59 \\
2017-01-02 & 57755.51 & \textit{i} & 0.68 & 26.51 \\
2017-01-02 & 57755.61 & \textit{g} & 0.69 & 26.75 \\
2017-01-21 & 57774.50 & \textit{z} & 0.52 & 26.31 \\ 
2017-01-23 & 57776.41 & \textit{r} & 0.83 & 26.44 \\
2017-01-23 & 57776.54 & \textit{i} & 0.70 & 26.43 \\
2017-01-25 & 57778.45 & \textit{g} & 1.77 & 26.13 \\
2017-01-26 & 57779.52 & \textit{z} & 0.73 & 24.36 \\
2017-01-30 & 57783.43 & \textit{i} & 0.74 & 26.03 \\
2017-01-30 & 57783.55 & \textit{z} & 0.65 & 25.85 \\
2017-02-01 & 57785.39 & \textit{g} & 0.66 & 26.38 \\
2017-02-02 & 57786.45 & \textit{r} & 0.65 & 26.60 \\
2017-02-02 & 57786.59 & \textit{i} & 0.49 & 25.83 \\
2017-02-21 & 57805.37 & \textit{z} & 0.64 & 25.69 \\
2017-02-23 & 57807.37 & \textit{g} & 1.40 & 26.30 \\
2017-02-23 & 57807.48 & \textit{r} & 0.91 & 26.33 \\
2017-02-25 & 57809.40 & \textit{i} & 0.75 & 25.85 \\
2017-03-04 & 57816.31 & \textit{z} & 0.64 & 25.73 \\
2017-03-04 & 57816.47 & \textit{i} & 0.67 & 26.33 \\
2017-03-06 & 57818.51 & \textit{r} & 0.73 & 26.47 \\
2017-03-22 & 57834.32 & \textit{g} & 0.84 & 26.74 \\
2017-03-22 & 57834.43 & \textit{z} & 0.56 & 25.82 \\
2017-03-23 & 57835.26 & \textit{i} & 0.67 & 25.89 \\
2017-03-25 & 57837.27 & \textit{r} & 0.98 & 26.11 \\
2017-03-29 & 57841.29 & \textit{g} & 0.92 & 26.53 \\
2017-03-29 & 57841.41 & \textit{z} & 0.74 & 25.60 \\
2017-03-30 & 57842.27 & \textit{i} & 0.98 & 26.02 \\
2017-04-01 & 57844.33 & \textit{r} & 1.18 & 26.13 \\
2017-04-23 & 57866.25 & \textit{r} & 0.94 & 26.10 \\
2017-04-23 & 57866.36 & \textit{z} & 0.81 & 25.32 \\
2017-04-26 & 57869.27 & \textit{i} & 1.25 & 25.70 \\
2017-04-26 & 57869.33 & \textit{g} & 0.88 & 26.45 \\
2017-04-27 & 57870.35 & \textit{i} & 0.55 & 26.09 \\
2017-04-29 & 57872.26 & \textit{z} & 0.74 & 25.30 \\
2017-06-20 & 57924.28 & \textit{z} & 1.15 & 23.95 \\
\hline
\multicolumn5c{\textbf{Season 2 (2018)}} \\
2018-01-10 & 58128.47 & \textit{i} & 1.12 & 26.02 \\
2018-01-10 & 58128.58 & \textit{z} & 1.87 & 25.04 \\
2018-01-13 & 58131.45 & \textit{i} & 1.13 & 25.96 \\
2018-01-13 & 58131.57 & \textit{z} & 1.23 & 25.74 \\
2018-02-09 & 58158.52 & \textit{i} & 1.58 & 25.64 \\
2018-02-09 & 58158.60 & \textit{z} & 2.03 & 24.50 \\
2018-02-10 & 58159.47 & \textit{r} & 1.27 & 26.51 \\
2018-02-10 & 58159.55 & \textit{z} & 1.19 & 25.58 \\
2018-03-10 & 58187.32 & \textit{i} & 0.94 & 25.26 \\
2018-03-10 & 58187.41 & \textit{z} & 0.71 & 24.73 \\
2018-03-18 & 58195.30 & \textit{z} & 0.75 & 26.12 \\
2018-03-18 & 58195.41 & \textit{i} & 0.81 & 26.26 \\
2018-04-21 & 58229.37 & \textit{i} & 1.50 & 23.09 \\
\hline
\multicolumn5c{\textbf{Season 3 (2018 - 2019)}} \\
2018-12-14 & 58466.56 & \textit{z} & 1.79 & 25.11 \\
2019-01-09 & 58492.42 & \textit{z} & 0.68 & 25.25 \\
2019-02-28 & 58542.34 & \textit{z} & 1.38 & 25.04 \\
2019-04-04 & 58577.32 & \textit{g} & 0.85 & 27.24 \\
2019-05-09 & 58612.31 & \textit{i} & 0.57 & 26.68 \\
\hline
\multicolumn5c{\textbf{Season 4 (2019 - 2020)}} \\
2019-10-26 & 58782.62 & \textit{r} & 1.14 & 22.93 \\
2019-12-01 & 58818.56 & \textit{i} & 0.99 & 22.68 \\
2019-12-31 & 58848.59 & \textit{z} & 1.80 & 24.68 \\
2020-01-20 & 58868.38 & \textit{i} & 0.92 & 25.60 \\
2020-02-20 & 58899.31 & \textit{i} & 0.74 & 26.39 \\
2020-02-20 & 58899.48 & \textit{z} & 0.75 & 25.97 \\
2020-02-23 & 58902.29 & \textit{g} & 0.90 & 26.75 \\
2020-02-23 & 58902.62 & \textit{r} & 1.13 & 25.80 \\
2020-02-26 & 58905.30 & \textit{z} & 0.73 & 25.59 \\
2020-02-28 & 58907.34 & \textit{r} & 1.26 & 26.28 \\
\enddata
%\tablecomments{
%To be precise, the reference images are taken with the \textit{r} and \textit{i} filters, while the \textit{r2} and \textit{i2} bands are used during the transient survey. The difference in these filters are small and we simply write \textit{r} and \textit{i}.
%}
\end{longdeluxetable}
\begin{longdeluxetable}{cccc}
\tablenum{A.2}
\tablecaption{
Photometry of HSC16aayt.
\label{tab:hsc16aaty}}
\tablewidth{0pt}
\tablehead{
\colhead{Filter} & \colhead{MJD} & \colhead{Magnitude} &
 \colhead{Uncertainty} 
}
%\decimalcolnumbers
\startdata
\textit{g}&57755.62 &$ 23.587 $&$  0.013 $\\
&57778.44 &$ 23.616 $&$  0.015 $\\
&57785.38 &$ 23.369 $&$  0.033 $\\
&57807.37 &$ 23.644 $&$  0.016 $\\
&57834.31 &$ 23.650 $&$  0.012 $\\
&57841.29 &$ 23.626 $&$  0.019 $\\
&57869.33 &$ 23.588 $&$  0.023 $\\
&58577.32 &$ 24.763 $&$  0.035 $\\
\hline
\textit{r}&57720.59 &$ 23.201 $&$  0.023 $\\
&57747.53 &$ 23.160 $&$  0.019 $\\
&57776.40 &$ 23.131 $&$  0.008 $\\
&57807.49 &$ 23.122 $&$  0.011 $\\
&57818.52 &$ 23.129 $&$  0.008 $\\
&57837.26 &$ 23.128 $&$  0.015 $\\
&57844.33 &$ 23.168 $&$  0.038 $\\
\hline
\textit{i}&57717.61 &$ 23.110 $&$  0.027 $\\
&57721.54 &$ 23.063 $&$  0.041 $\\
&57747.61 &$ 22.961 $&$  0.016 $\\
&57755.55 &$ 22.982 $&$  0.014 $\\
&57776.53 &$ 22.977 $&$  0.009 $\\
&57783.44 &$ 22.867 $&$  0.023 $\\
&57786.59 &$ 22.939 $&$  0.011 $\\
&57809.41 &$ 22.915 $&$  0.015 $\\
&57816.48 &$ 22.879 $&$  0.010 $\\
&57835.25 &$ 22.874 $&$  0.015 $\\
&58195.40 &$ 23.147 $&$  0.022 $\\
&58229.39 & $> 22.328 $& \nodata \\
&58612.31 &$ 24.322 $&$  0.029 $\\
\hline
\textit{z}&57715.53 &$ 23.363 $&$  0.083 $\\
&57721.59 &$ 23.088 $&$  0.035 $\\
&57745.56 &$ 22.991 $&$  0.023 $\\
&57755.45 &$ 22.963 $&$  0.051 $\\
&57774.49 &$ 22.871 $&$  0.010 $\\
&57779.53 &$ 22.917 $&$  0.043 $\\
&57783.55 &$ 22.863 $&$  0.014 $\\
&57805.37 &$ 22.837 $&$  0.021 $\\
&57816.30 &$ 22.854 $&$  0.027 $\\
&57834.45 &$ 22.801 $&$  0.021 $\\
&57841.40 &$ 22.742 $&$  0.029 $\\
&58128.60 &$ 22.758 $&$  0.041 $\\
&58131.58 &$ 22.860 $&$  0.025 $\\
&58158.66 &$ 22.507 $&$  0.121 $\\
&58159.55 &$ 23.102 $&$  0.034 $\\
&58195.28 &$ 23.199 $&$  0.033 $\\
&58466.57 &$ 23.456 $&$  0.105 $\\
&58492.41 &$ 24.016 $&$  0.106 $\\
&58542.34 &$ 23.926 $&$  0.098 $\\
&58848.61 & $> 23.872 $& \nodata \\
&58905.32 & $> 24.838 $& \nodata \\
\enddata
%\tablecomments{
%To be precise, the reference images are taken with the \textit{r} and \textit{i} filters, while the \textit{r2} and \textit{i2} bands are used during the transient survey. The difference in these filters are small and we simply write \textit{r} and \textit{i}.
%}
\end{longdeluxetable}
\begin{longdeluxetable}{cccc}
\tablenum{A.3}
\tablecaption{
Photometry of HSC19edgb.
\label{tab:hsc19edgb}}
\tablewidth{0pt}
\tablehead{
\colhead{Filter} & \colhead{MJD} & \colhead{Magnitude} &
 \colhead{Uncertainty} 
}
%\decimalcolnumbers
\startdata
\textit{g}&57717.57 & $> 25.818 $& \nodata \\
&57755.62 & $> 27.277 $& \nodata \\
&57778.45 & $> 26.973 $& \nodata \\
&57785.39 & $> 26.793 $& \nodata \\
&57807.37 & $> 26.887 $& \nodata \\
&57834.32 & $> 27.139 $& \nodata \\
&57841.29 & $> 26.719 $& \nodata \\
&57869.33 & $> 26.427 $& \nodata \\
&58577.32 &$ 24.295 $&$  0.010 $\\
&58902.29 & $> 27.096 $& \nodata \\
\hline
\textit{r}&57720.60 & $> 26.707 $& \nodata \\
&57747.53 & $> 26.292 $& \nodata \\
&57776.41 & $> 26.906 $& \nodata \\
&57786.45 & $> 27.011 $& \nodata \\
&57807.48 & $> 26.807 $& \nodata \\
&57818.51 & $> 26.984 $& \nodata \\
&57837.26 & $> 26.466 $& \nodata \\
&57844.33 & $> 26.521 $& \nodata \\
&57866.25 & $> 26.077 $& \nodata \\
&58159.47 & $> 26.921 $& \nodata \\
&58782.62 & $> 23.015 $& \nodata \\
&58902.62 & $> 25.700 $& \nodata \\
&58907.34 & $> 26.853 $& \nodata \\
\hline
\textit{i}&57717.62 & $> 26.171 $& \nodata \\
&57721.54 & $> 26.012 $& \nodata \\
&57747.62 & $> 26.098 $& \nodata \\
&57755.52 & $> 26.789 $& \nodata \\
&57776.55 & $> 26.699 $& \nodata \\
&57783.43 & $> 26.338 $& \nodata \\
&57786.59 & $> 26.453 $& \nodata \\
&57809.41 & $> 26.297 $& \nodata \\
&57816.47 & $> 26.682 $& \nodata \\
&57835.26 & $> 26.286 $& \nodata \\
&57842.27 & $> 26.306 $& \nodata \\
&57869.27 & $> 25.979 $& \nodata \\
&57870.35 & $> 26.164 $& \nodata \\
&58128.47 & $> 26.292 $& \nodata \\
&58131.46 & $> 26.221 $& \nodata \\
&58158.52 & $> 26.003 $& \nodata \\
&58187.32 & $> 25.484 $& \nodata \\
&58195.41 & $> 26.560 $& \nodata \\
&58229.38 & $> 23.258 $& \nodata \\
&58612.32 &$ 23.115 $&$  0.007 $\\
&58818.56 & $> 22.759 $& \nodata \\
&58868.38 &$ 25.478 $&$  0.180 $\\
&58899.31 &$ 26.417 $&$  0.193 $\\
\hline
\textit{z}&57715.55 & $> 25.556 $& \nodata \\
&57721.60 & $> 25.670 $& \nodata \\
&57745.56 & $> 25.813 $& \nodata \\
&57755.45 & $> 25.727 $& \nodata \\
&57774.50 & $> 26.758 $& \nodata \\
&57779.53 & $> 24.744 $& \nodata \\
&57783.55 & $> 26.202 $& \nodata \\
&57805.37 & $> 26.020 $& \nodata \\
&57816.30 & $> 25.956 $& \nodata \\
&57834.44 & $> 26.182 $& \nodata \\
&57841.41 & $> 25.555 $& \nodata \\
&57866.36 & $> 25.442 $& \nodata \\
&57872.26 & $> 25.380 $& \nodata \\
&57924.28 & $> 23.965 $& \nodata \\
&58128.59 & $> 25.811 $& \nodata \\
&58131.57 & $> 26.312 $& \nodata \\
&58158.62 & $> 24.609 $& \nodata \\
&58159.55 & $> 25.904 $& \nodata \\
&58187.41 & $> 24.475 $& \nodata \\
&58195.30 & $> 26.269 $& \nodata \\
&58466.56 & $> 25.641 $& \nodata \\
&58492.42 &$ 22.462 $&$  0.014 $\\
&58542.34 &$ 22.655 $&$  0.012 $\\
&58848.60 &$ 24.685 $&$  0.136 $\\
&58899.48 &$ 25.783 $&$  0.149 $\\
&58905.31 & $> 25.772 $& \nodata \\
\enddata
%\tablecomments{
%To be precise, the reference images are taken with the \textit{r} and \textit{i} filters, while the \textit{r2} and \textit{i2} bands are used during the transient survey. The difference in these filters are small and we simply write \textit{r} and \textit{i}.
%}
\end{longdeluxetable}
\begin{longdeluxetable}{cccc}
\tablenum{A.4}
\tablecaption{
Photometry of HSC19edge.
\label{tab:hsc19edge}}
\tablewidth{0pt}
\tablehead{
\colhead{Filter} & \colhead{MJD} & \colhead{Magnitude} &
 \colhead{Uncertainty} 
}
%\decimalcolnumbers
\startdata
\textit{g}&57717.56 & $> 25.649 $& \nodata \\
&57755.61 & $> 26.863 $& \nodata \\
&57778.45 & $> 26.863 $& \nodata \\
&57785.39 & $> 26.766 $& \nodata \\
&57807.37 & $> 26.825 $& \nodata \\
&57834.32 & $> 26.995 $& \nodata \\
&57841.29 & $> 26.631 $& \nodata \\
&57869.33 & $> 26.326 $& \nodata \\
&58577.32 &$ 25.989 $&$  0.054 $\\
&58902.29 & $> 27.067 $& \nodata\\
\hline
\textit{r}&57720.60 & $> 26.353 $& \nodata \\
&57747.53 & $> 26.117 $& \nodata \\
&57776.41 & $> 26.728 $& \nodata \\
&57786.45 & $> 26.724 $& \nodata \\
&57807.48 & $> 26.326 $& \nodata \\
&57818.52 & $> 26.427 $& \nodata \\
&57837.26 & $> 26.460 $& \nodata \\
&57844.33 & $> 26.450 $& \nodata \\
&57866.25 & $> 26.047 $& \nodata \\
&58159.47 & $> 26.715 $& \nodata \\
&58782.62 & $> 22.923 $& \nodata \\
&58902.62 & $> 25.884 $& \nodata \\
&58907.34 & $> 26.762 $& \nodata \\
\hline
\textit{i}&57717.62 & $> 26.137 $& \nodata \\
&57721.55 & $> 25.948 $& \nodata \\
&57747.62 & $> 25.988 $& \nodata \\
&57755.52 & $> 26.414 $& \nodata \\
&57776.54 & $> 26.654 $& \nodata \\
&57783.43 & $> 26.223 $& \nodata \\
&57786.59 & $> 26.329 $& \nodata \\
&57809.41 & $> 26.134 $& \nodata \\
&57816.47 & $> 26.549 $& \nodata \\
&57835.26 & $> 26.184 $& \nodata \\
&57842.27 & $> 26.218 $& \nodata \\
&57869.27 & $> 25.825 $& \nodata \\
&57870.35 & $> 25.942 $& \nodata \\
&58128.47 & $> 26.137 $& \nodata \\
&58131.45 & $> 26.061 $& \nodata \\
&58158.52 & $> 25.863 $& \nodata \\
&58187.32 & $> 25.301 $& \nodata \\
&58195.41 & $> 26.280 $& \nodata \\
&58229.37 & $> 22.964 $& \nodata \\
&58612.31 &$ 25.663 $&$  0.067 $\\
&58818.56 & $> 22.735 $& \nodata \\
&58868.38 & $> 25.694 $& \nodata \\
&58899.31 &$ 26.251 $&$  0.164 $\\
\hline
\textit{z}&57715.54 & $> 25.520 $& \nodata \\
&57721.60 & $> 25.415 $& \nodata \\
&57745.56 & $> 25.791 $& \nodata \\
&57755.45 & $> 25.659 $& \nodata \\
&57774.50 & $> 26.521 $& \nodata \\
&57779.53 & $> 24.744 $& \nodata \\
&57783.55 & $> 26.108 $& \nodata \\
&57805.37 & $> 25.800 $& \nodata \\
&57816.31 & $> 25.759 $& \nodata \\
&57834.43 & $> 26.012 $& \nodata \\
&57841.41 & $> 25.665 $& \nodata \\
&57866.36 & $> 25.272 $& \nodata \\
&57872.27 & $> 25.182 $& \nodata \\
&57924.28 & $> 23.962 $& \nodata \\
&58128.58 & $> 25.591 $& \nodata \\
&58131.57 & $> 26.182 $& \nodata \\
&58158.60 & $> 24.544 $& \nodata \\
&58159.55 & $> 25.878 $& \nodata \\
&58187.41 & $> 24.718 $& \nodata \\
&58195.30 & $> 26.174 $& \nodata \\
&58466.55 &$ 23.264 $&$  0.030 $\\
&58492.42 &$ 23.244 $&$  0.035 $\\
&58542.35 &$ 23.189 $&$  0.023 $\\
&58848.59 &$ 23.495 $&$  0.054 $\\
&58899.48 &$ 23.820 $&$  0.025 $\\
&58905.30 &$ 23.927 $&$  0.050 $\\
\enddata
%\tablecomments{
%To be precise, the reference images are taken with the \textit{r} and \textit{i} filters, while the \textit{r2} and \textit{i2} bands are used during the transient survey. The difference in these filters are small and we simply write \textit{r} and \textit{i}.
%}
\end{longdeluxetable}

\newpage


\bibliography{sample63}{}
\bibliographystyle{aasjournal}



\end{document}